\documentclass[preprint,authoryear]{elsarticle}
\usepackage{amsmath, amssymb}
\usepackage{hyperref}
\usepackage{graphicx}
\usepackage{color}
\usepackage{natbib}
\usepackage{booktabs}
\textfloatsep = 0.4in \addtocontents{toc}{\vspace{0.4in} \hfill
Page\endgraf} \addtocontents{lof}{\vspace{0.2in} \hspace{0.13in} \
Figure\hfill Page\endgraf} \addtocontents{lot}{\vspace{0.2in}
\hspace{0.13in} \ Table\hfill Page\endgraf}
\usepackage{float}
\usepackage{textcomp,bm}
\usepackage{pdfpages}
\usepackage{array}
\usepackage{rotating}
\usepackage{listings}
\usepackage{pdflscape}
\usepackage{setspace}
\usepackage{mathptmx}
\usepackage{capt-of}
\usepackage[table, svgnames]{xcolor}
\usepackage{colortbl}
\usepackage{subfig}
\usepackage{wrapfig}
\usepackage{times}
\usepackage{float}
\usepackage{bm}
\usepackage{lipsum}
\usepackage{url}
\usepackage{multirow}
\usepackage{pdfpages,caption}
\usepackage[subfigure, titles]{tocloft}
\usepackage{caption}
\usepackage{datetime}
\usepackage{algorithm}
\usepackage{cite}
\usepackage{lscape}
\usepackage{mdframed}
\usepackage{array}
\usepackage{tikz}
\usetikzlibrary{bayesnet,positioning}
\usepackage{makecell}
\usepackage{titletoc}
\usepackage{tikz}
\usetikzlibrary{decorations.pathmorphing} 
\usetikzlibrary{fit}
\usetikzlibrary{backgrounds}	
\usetikzlibrary{positioning}
\usetikzlibrary{shapes,decorations,arrows,calc,arrows.meta,fit,positioning}
\tikzset{
    -Latex,auto,node distance =1 cm and 1 cm, semithick,
    state/.style ={ellipse, draw, minimum width = 0.7 cm},
    point/.style = {circle, draw, inner sep=0.04cm,fill,node contents={}},
    bidirected/.style={Latex-Latex,dashed},
    el/.style = {inner sep=2pt, align=left, sloped}
     latentnode/.style={draw, minimum width=5mm, shape=circle, ultra thick, black},
  dagconn/.style={arrows=->, black, thick},
  plate/.style={draw, shape=rectangle, rounded corners=0.5ex, thick, minimum width=3.1cm, text width=3.1cm,inner sep=10pt, inner ysep=10pt, 
    label={[xshift=-14pt,yshift=14pt]south east:#1}}
}

\usepackage[nottoc]{tocbibind}
\IfFileExists{upquote.sty}{\usepackage{upquote}}{}

\begin{document}
\pagenumbering{alph}
\begin{frontmatter}
\title{A Bayesian Spatio-Temporal Top-Down Framework for Estimating Opioid Use Disorder Risk Under Data Sparsity}

\title{A Bayesian Spatio-Temporal Top-Down Framework for Estimating Opioid Use Disorder Risk Under Data Sparsity}

\author[biostats]{Emily N. Peterson\corref{cor1}}
\ead{emily.n.peterson@emory.edu}

\author[biostats]{Alex A. Edwards}
\author[policy]{Martha Wetzel}
\author[biostats]{Lance A. Waller}
\author[behavior]{Hannah Cooper}
\author[policy]{Courtney Yarbrough}

\cortext[cor1]{Corresponding author}

\address[biostats]{Department of Biostatistics and Bioinformatics, Emory Rollins School of Public Health, Atlanta, GA, USA}
\address[policy]{Department of Health Policy and Management, Emory Rollins School of Public Health, Atlanta, GA, USA}
\address[behavior]{Department of Behavioral Sciences and Health Education, Emory Rollins School of Public Health, Atlanta, GA, USA}

\begin{abstract}
County-level estimates of opioid use disorder (OUD) are essential for understanding the influence of local economic and social conditions. They provide policymakers with the granular information needed to identify, target, and implement effective interventions and allocate resources appropriately. Traditional disease mapping methods typically rely on Poisson regression, modeling observed counts while adjusting for local covariates that are treated as fixed and known. However, these methods may fail to capture the complexities and uncertainties in areas with sparse or absent data. To address this challenge, we developed a Bayesian hierarchical spatio-temporal top-down approach designed to estimate county-level OUD rates when direct small-area (county) data is unavailable. This method allows us to infer small-area OUD rates and quantify associated uncertainties, even in data-sparse environments using observed state-level OUD rates and a combination of state and county level informative covariates. We applied our approach to estimate OUD rates for 3,143 counties in the United States between 2010 and 2025. Model performance was assessed through simulation studies.
\end{abstract}

\end{frontmatter}

\pagenumbering{roman} \setcounter{page}{1}
\captionsetup{font=small,skip=0pt}


\normalsize

\pagenumbering{arabic}
\setcounter{page}{1}
\singlespacing

\section{Introduction}
Monitoring opioid use disorder (OUD) remains a pressing public health priority that necessitates robust surveillance systems capable of tracking trends, identifying high-risk populations, and guiding targeted interventions. In 2021, an estimated 2.5 million adults aged 18 years and older were affected by OUD in the United States \citep{NIDA_OUD_2023, NIDA_OUD_2023A}. That same year, approximately 80,411 overdose deaths were attributed to opioids, representing 75.4\% of all drug overdose fatalities \citep{opioid_overdose}. The National Survey on Drug Use and Health (NSDUH) plays a central role in providing national estimates of OUD prevalence, offering critical insights into treatment access and demographic patterns \citep{NSDUH_OUD_2023}. Despite these efforts, a substantial treatment gap persists, underscoring the need for improved data infrastructure. The U.S. Department of Health and Human Services (HHS) has emphasized the importance of comprehensive data collection strategies to close key knowledge gaps and inform policy responses to the opioid epidemic \citep{HHS_Opioid_2020, HHS_Opioid_2020A}. A major challenge to these efforts is that OUD prevalence data are only available at the state-level and only for selected years, leaving a critical void in our understanding of finer-scale patterns. While indirect indicators—such as opioid-related mortality and treatment admissions—can serve as proxies, these sources are often biased due to the illicit nature of drug use and underreporting in self-reported surveys. Consequently, current surveillance systems fall short of capturing the true burden of OUD at the local level \citep{hepler_integrated_2023, kline2021multivariate, kline_estimating_2021, kline_dynamic_2023}. Given the well-documented heterogeneity in the social and structural determinants of substance use—including poverty, healthcare access, housing instability, and racialized policing—more granular estimates are urgently needed to inform place-based interventions \citep{Dasgupta2018,  Monnat2018,  Rosenblum2022}. To address this significant data gap, we introduce a Bayesian hierarchical spatio-temporal top-down framework to generate robust, small-area estimates of OUD at the county level across the United States from 2010 to 2025. This top-down modeling approach integrates multiple data sources and captures spatial and temporal dependencies to produce granular, policy-relevant estimates. By offering a more accurate and localized view of the opioid crisis, our methodology supports data-driven public health responses and enhances the precision of resource allocation.\\

Bayesian small area estimation (SAE) methods allow for sharing of information across spatially structured neighboring regions thereby reducing uncertainty surrounding small areas estimates of health indicators \citep{havard_rue_gaussian_2005, waller_gotway, jon_wakefield_disease_2007}. However, one of the most pervasive challenges in SAE arises when small areas lack directly observed data, such as when subpopulations of interest are sparsely populated or suffer from an absence of data \citep{ghosh1994small, rao2015small}. In these cases, hierarchical Bayesian models commonly rely on indirect estimates by borrowing strength and information from larger aggregates (e.g., state- or national-level observed data) and distributing these aggregates across small areas (e.g., latent county estimates) based on covariates, population-at-risk size, and hierarchical population structures (e.g., commonly assume that counts of health indicators of interest are consistent in that they aggregate up to larger geographic areas, i.e., county level counts aggregate up to state-level counts) \citep{ghosh1994small, rao2015small,  census2020}. Estimating latent (unknown) risks for smaller geographic areas given reported data from larger aggregated geographic areas is known as a process of downscaling \citep{arambepola_simulation_2022, griffin_spatial_2020,  nandi_disaggregation_2020, python_downscaling_2022}. Previous studies have applied a downscaling approach to produce fine-resolution spatial estimates by leveraging coarser-scale data and incorporating auxiliary information to infer patterns at smaller geographic units. \citet{linard_large-scale_2012} derived high resolution gridded population-at-risk estimates using population estimates at larger spatial scales. \citet{griffin_spatial_2020} developed a spatial downscaling disease risk estimation model using random forests machine learning. \citet{arambepola_simulation_2022} and \citep{nandi_disaggregation_2020} assessed robustness of disaggregation regression using simulation studies. \citet{konstantinoudis_long-term_2021} assessed localized patterns of air pollution exposure and COVID-19 mortality at the Lower Layer Super Output Area (LSOA) level—very fine spatial resolution using exposure data derived from coarser-resolution pollution models. \citet{python_downscaling_2022} compared counts of COVID-19 cases in China at district level  with fine spatial scale predictions from a Bayesian downscaling regression model applied to province level data. While the broader literature on Bayesian hierarchical downscaling models is rich, few studies explicitly address the issue of downscaling using sparse and biased data at larger spatial levels. Recent methodological work has begun to address overdispersion \citep{wulandari2023overdispersion}, benchmark constraints \citep{Ugarte2020benchmarking}, and spatial misalignment \citep{Bradley2015multiscale}, but further development is needed to ensure interpretability and stability of disaggregated estimates in highly sparse and biased settings. We propose a two-stage modular downscaling approach to reduce instabilities in small area estimates in the context of sparse or biased data at aggregated levels. \\

In this paper, we develop a two-stage Bayesian hierarchical spatio-temporal framework to estimate state and county-level OUD risk from observed state-level counts and informative covariates. Our top-down disaggregation model introduces several innovations. First, we incorporate a population-weighted scaling mechanism that stabilizes estimates in sparsely populated areas and prevents distortion from extreme scaling factors. Scaling factors serve as weights that modulate the contribution of aggregate data to small area estimates, often in conjunction with covariates or spatial predictors. They are intended to provide a realistic distribution of aggregate counts, proportional to the size and characteristics of each smaller area. For instance, a sparsely populated county would typically receive a smaller share of state-level counts compared to a densely populated county, assuming other factors are equal. A specific and under-addressed issue in these models is scaling instability, which occurs when small area populations are disproportionately small compared to the larger aggregates. When population sizes vary significantly and population sizes for some areas are extremely small, the scaling factors can become extreme. These disproportionate values can dominate the model's output, leading to implausible estimates that fail to accurately reflect the true underlying distribution of the health indicator. This problem is compounded when covariates or predictors fail to sufficiently moderate these scaling effects, as often happens in sparse or extreme data settings \citep{gao2022spatial, hogg2023two, mercer2015small, peterson2023bayesian}. The failure to address this scaling issue has significant implications for the reliability of small area estimates. We introduce a population-weighted correction to mitigate scaling instability in hierarchical models. Second, we explicitly model changes in diagnostic definitions over time via an adjustment factor that ensures temporal consistency of defined cases and aims to model the overall increasing trend in county-level OUD rates. Third, we explicitly quantify and propagate uncertainty across modeling stages using a fully Bayesian framework. By adopting a sequential (modular) approach, we ensure that state-level estimates are informed only by state-level covariates and assumptions, without contamination from downstream county-level information and uncertainty. This modularization preserves the integrity of the top-down structure and prevents feedback loops that can distort posterior distributions. As shown in prior work, modular Bayesian inference, often implemented through cut models, provides a principled strategy to avoid spurious correlations and overfitting in hierarchical systems \citep{bennett_errors--variables_2001, jacob_better_2017, peterson2024bayesian, plummer_cuts_2015}. Finally, we validate our model through a simulation study and apply it to estimate county-level OUD burden in the U.S. from 2010-2025.\\

The remainder of this paper is organized as follows. Section \ref{sec:data} describes the data used to inform county and state-level estimates. Section~\ref{sec:methods} describes the hierarchical modeling framework and adjustment mechanism. Section~\ref{sec:simulation} presents the simulation design and model validation. Section~\ref{sec:results} illustrates both state and county model results, and findings of the simulation study. Section~\ref{sec:discussion} concludes with policy implications, model limitations, and directions for future research.

\section{Data}\label{sec:data}
\begin{figure}[H]
    \centering
    \includegraphics[width=0.6\textwidth]{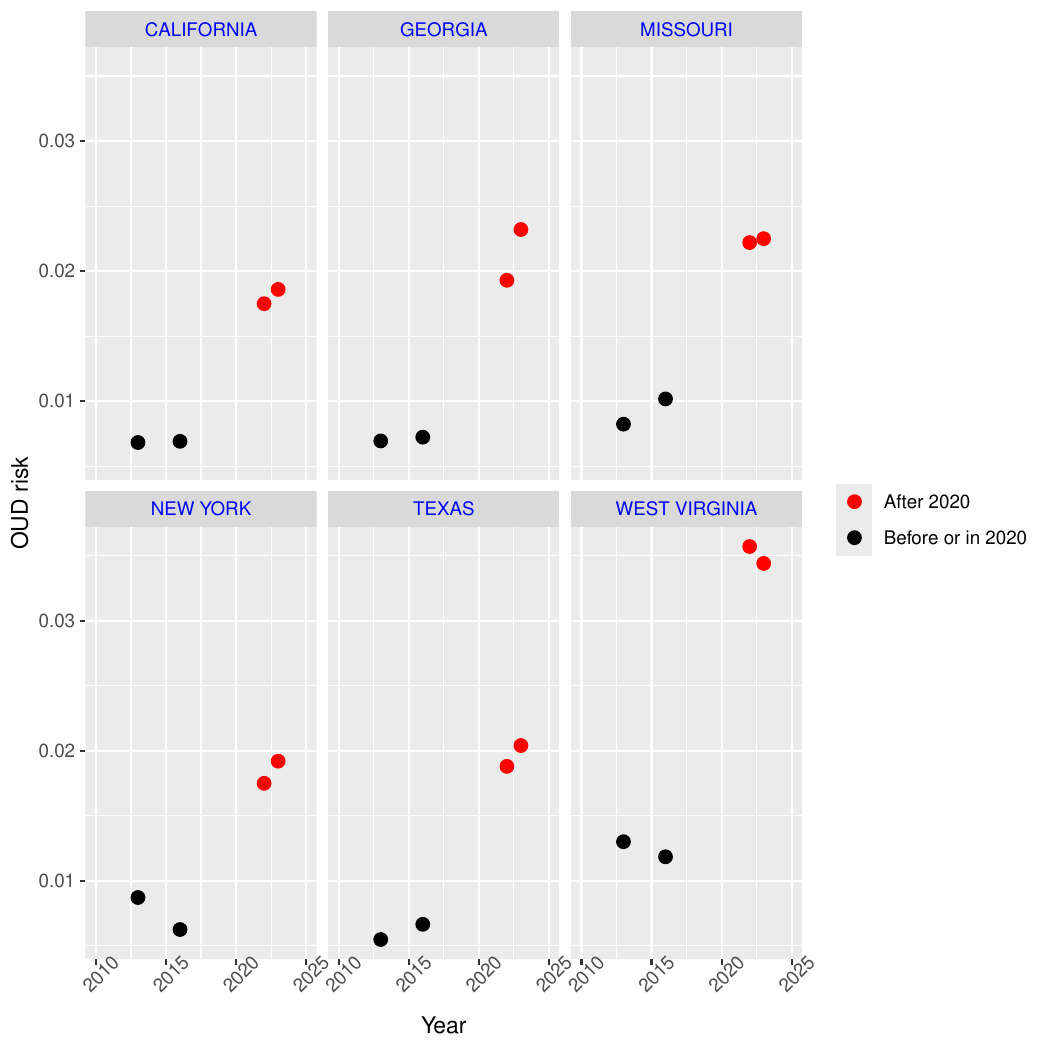}
    \caption{NSDUH reported OUD risks (counts/population at risk) for selected states. The black dots indicate OUD counts out of the population of respondents who reported misusing opioids. Red dots indicate OUD counts out of the population of respondents who reported any prescription opioid use.}
    \label{fig:state_oud_trends}
\end{figure}

Our analysis integrates multiple data sources to inform estimation of OUD incidence at both the state and county levels. State-level OUD prevalence data were obtained from the National Survey on Drug Use and Health (NSDUH), which provides estimates of substance use for individuals aged 12 and older \citep{NSDUH2014, NSDUH2022}. Population denominators of the 12+ years old populations at the state and county levels were derived from the U.S. Census Bureau's Population Estimates Program (PEP), providing consistent yearly counts from 2010 to 2023 \citep{pep, USCensusPEP}. To inform spatial and temporal variation in risk, we incorporated state-level informative covariates including rates of past-year pain reliever misuse (PR misuse) and heroin use from NSDUH. At the county-level, we incorporated informative covariates of: (1) mortality rates from any opioid use, (2) percent rural population, (3) poverty rate, (4) disability, and (5) cumulative opioid rate from 2011. County-level covariates were harmonized across years and standardized prior to modeling. Refer to Appendix \ref{sec:covs} for a complete set of graphical representations of state and county level covariate trends. Figure \ref{fig:state_oud_trends} illustrates observed data for selected states. Prior to 2020, the NSDUH assessed opioid use disorder (OUD) only among respondents who reported misusing opioids (black dots). Starting in 2021, the NSDUH expanded its methodology to assess OUD among all respondents who reported any prescription opioid use, regardless of whether they reported misuse (red dots). Due to methodological changes implemented in 2020, including the adoption of DSM-5 criteria and alterations in data collection procedures, the counts of individuals identified with opioid use disorder (OUD) are not directly comparable between years prior to 2020 and those from 2020 onward. These changes have impacted the way OUD is assessed and reported, leading to differences in prevalence estimates that reflect both actual trends and modifications in survey methodology \citep{CBHSQ2021}. In this study, we obtain estimates of OUD prevalence using the OUD definition prior to 2020, which reflect the number of individuals with OUD out of respondents who misuse opioids. As such, we incorporate definition-related adjustment factors further described in Section \ref{sec:methods}.

\section{Methods}\label{sec:methods}

\subsection{Overview}
\begin{figure}[ht]
\centering
\begin{tikzpicture}[node distance=1.2cm and 1.6cm]

\node[latent] (lambda) {$\lambda_{s,t}$};
\node[latent, above left=of lambda] (alpha) {$\alpha$};
\node[latent, above=of lambda] (beta1) {$\beta_1$};
\node[latent, above right=of lambda] (beta2) {$\beta_2$};
\node[latent, below=of lambda] (rate) {$\pi_{s,t}$};
\node[latent] (delta) at ($(rate) + (-2,-0.5)$) {$\Delta_{s,t}$};
\node[latent, , fill=gray!30, left=0.5cm of delta] (r) {$r_{s,t}$};
\node[latent, fill=gray!30, below=of rate] (yi) {$y_i$};
\node[latent, fill=gray!30, left=of yi] (denom) {$n_i$};
\node[latent, fill=gray!30, below left=of lambda] (x) {$\mathbf{x}_{s,t}$};
\node[latent, left= of lambda] (phi) {$\phi_{s,t}$};
\node[latent, right=of lambda] (omega) {$\omega_s$};

\edge {alpha,beta1,beta2} {lambda};
\edge {x} {lambda};
\edge {phi} {lambda};
\edge {omega} {lambda};
\edge {lambda} {rate};
\edge {rate} {yi};
\edge {r} {delta};
\edge {delta} {yi};
\edge {denom} {yi};

\node[plate, inner sep=20pt, draw=blue, thick,
      fit= (alpha) (beta1) (beta2) (lambda) (x) (delta) (r) (rate) (yi) (denom) (phi) (omega),
      label={[yshift=1em]above:{\textbf{Stage I}}}] (plate1) {};

\node[latent, right=2cm of omega,  draw=none] (omtilde) {$\tilde{\omega}_{s}$};
\node[latent, right=3cm of rate,  draw=none] (pitilde) {$\tilde{\pi}_{s,t} \cdot N_{s,t} = \tilde{y}_{s,t}$};

\node[draw=black, dashed, inner sep=5pt, fit={(omtilde)(pitilde)}, 
      label={[yshift=1em]above:{\textbf{Post-Processing}}}, 
      name=plate3] {};

\node[latent, right=8cm of lambda] (eta) {$\eta_{c,t}$};
\node[latent, above left=of eta] (alpha2) {$\alpha$};
\node[latent, fill=gray!30, right=of eta] (xc) {$\mathbf{x}_{c,t}$};
\node[latent, above=of eta] (betas) {$\beta_{1:5}$};
\node[latent, above right=of eta] (fcounty) {$\phi_{c,t}$};
\node[latent, below=of eta] (mu) {$\rho_{c,t}$};
\node[latent, below=of mu] (yc) {$\mu_{c,t}$};
\node[plate, inner sep=20pt, draw=green, thick,
      fit= (alpha2) (betas) (eta) (xc) (mu) (yc) (fcounty),
      label={[yshift=1em]above:{\textbf{Stage II}}}] (plate2) {};

\edge {alpha2,betas,fcounty} {eta};
\edge {eta} {mu};
\edge {mu} {yc};
\edge {xc} {eta};

\draw[->, dashed, thick] (rate) -- (pitilde);
\draw[->, dashed, thick] (pitilde) -- (yc);
\draw[->, dashed, thick] (omega) -- (omtilde);
\draw[->, dashed, thick] (omtilde) -- (eta);

\end{tikzpicture}
\caption{Directed acyclic graph (DAG) illustrating the two-stage hierarchical Bayesian top-down model approach. Observed data quantities are denoted with gray shaded circles. Latent nodes are denoted with clear circles. The post-processed inputs obtained from Stage I are used in Stage II and are denoted with $\tilde{x}$.}
\label{fig:dag}
\end{figure}
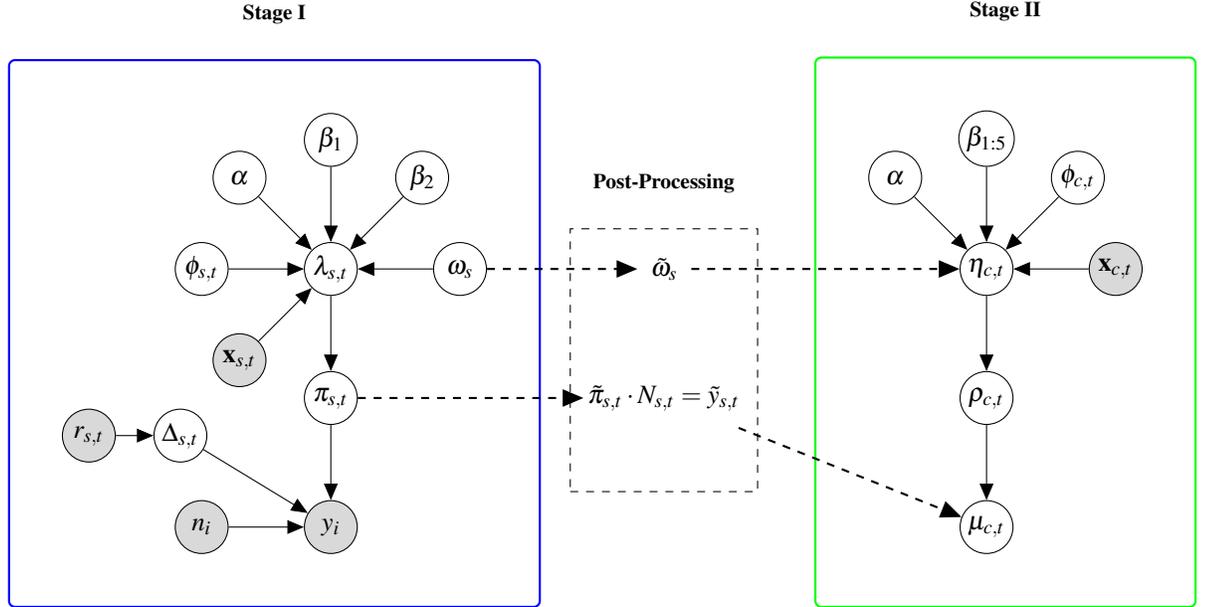

We develop a two-stage Bayesian spatial top-down estimation (B-Step) modeling framework to estimate OUD risk (count/population-at-risk) at state and county levels in the context of data sparsity. Figure \ref{fig:dag} illustrates this two stage approach using a directed acyclic graph. A full summary of notation is given in Appendix \ref{sec:A}. The first stage (blue box) leverages state-level surveillance data $y_i$ to model state-level OUD risk for state $s$ and year $t$, $\pi_{s,t}$, using a hierarchical Poisson time series regression model including state-year specific covariates $\mathbf{X}_{s,t}$, a state specific intercept $\omega_s$, and a state-specific temporal process $\phi_{s,t}$. We incorporate an adjustment term $r_{s,t}$ that adjusts for changes to OUD case definitions for years 2020 and onward capturing the switch to the DSM-5 criteria. In Stage II (green box), predictive posterior distributions of state-level OUD counts $\mathbf{\tilde{y}}_{s,t}$ and state-level intercepts $\mathbf{\tilde{\omega}}_s$ are used as inputs to the county-specific model to ensure consistency of downscale disaggregation to county levels counts from state-level totals. Latent county-year OUD risks are modeled assuming a spatio-temporal latent process structure, and are informed by known county-year specific covariates $\mathbf{x}_{c,t}$. This approach accounts for differential data availability, borrowing strength across space and time, and yields fully probabilistic estimates with quantified uncertainty.

\subsection{Stage 1: State-Level Bayesian Poisson Model}
In stage I, we aim to generate stable estimates of OUD rates at the state-level from years 2010 to 2025 where surveillance data is sparse. We specify a Bayesian hierarchical Poisson regression model that estimates state-year-specific OUD risk while accounting for population denominators, changes in diagnostic definitions over time, and structured state-level temporal variation. The model incorporates fixed effects for time-varying covariates, state-specific random intercepts to account for persistent spatial heterogeneity, and temporally structured random effects to capture evolving state-level trends. 

\paragraph*{Data Model}
Let $y_i$ denote the number of observed OUD cases for the $i^{th}$ observation corresponding to state-year $(s[i], t[i])$ with known population at risk $n_i$ (population of ages 12+). We model $y_i$ assuming a Poisson data model given in Eq. \ref{eq:stdm}:
\begin{equation}\label{eq:stdm}
    y_i |\pi_{s,t}, n_i, r_{s,t}\sim \text{Poisson}\left( \frac{n_i \cdot \pi_{s[i],t[i]}}{\Delta_{s[i],t[i]}} \right)
\end{equation}
where $\pi_{s,t}$ is the latent true OUD incidence risk for state $s$ and year $t$.

\paragraph*{Latent Process Model}
We model the log risk for state $s$ and year $t$ $\pi_{s,t}$ as:
\begin{equation}
    \log(\pi_{s,t}) = \lambda_{s,t} = \alpha + \beta_1 x_{1,s,t} + \beta_2 x_{2,s,t} + \phi_{s,t} + \omega_s
\end{equation}
where $x_{1,s,t}$ and $x_{2,s,t}$ are state-year specific rates of PR misuse and heroin use , $\omega_s$ is a state-level random effect accounting for time-invariant state-specific heterogeneity which is modeled assuming a $\omega_s \sim \mathcal{N}(0, 10)$, and $\phi_{s,t}$ is a state-specific temporally-structured deviation term that accounts for temporal trends within states modeled via a first order random walk over years $t=1, ...,T$ with reference year $t_{\text{ref}} = 2015$ given by Eq. \ref{eq:rw}.
\begin{align}\label{eq:rw}
    \phi_{s,t_{\text{ref}}} &\sim \mathcal{N}(0, \tau_\phi^{-1}) \\
    \phi_{s,t} &\sim \mathcal{N}(\phi_{s,t-1}, \tau_\phi^{-1}), \quad t > t_{\text{ref}} \nonumber\\
    \phi_{s,t-1} &\sim \mathcal{N}(\phi_{s,t}, \tau_\phi^{-1}), \quad t < t_{\text{ref}}\nonumber
\end{align}

\paragraph*{Adjustment Factors}
\noindent
 We incorporate a model-based adjustment factor, denoted $\Delta_{s,t}$, to account for changes in diagnostic criteria that affect the comparability of observed OUD case counts over time. Specifically, prior to $t_0 = 2020$, the reported OUD numerator included only individuals who both reported opioid use and were subsequently screened for OUD under diagnostic criteria from DSM-IV. Beginning in 2020, the survey protocol changed such that all individuals reporting any opioid use were screened for OUD under the definition from the DSM-V, resulting in a broader case definition under the and a corresponding increase in observed OUD rates. To correct for this shift, we computed $r_{s,t}$ as the ratio of observed OUD case counts in 2016 to the corresponding count in year $t$ for each state. In summary, $r= y_{s,2016} / y_{s,t}$, where $y_{s,t}$ is the reported count in state $s$ and year $t$. For years prior to 2020, we fixed $r_{s,t} = 1$, i.e, no adjustment. If a direct comparison year was missing, e.g., for years 2017 to 2020, we linearly interpolated the missing $r_{s,t}$ values assuming an increasing trend in state-level adjusted counts. This approach ensures temporal comparability of OUD rates despite structural changes in surveillance practices. Eq. \ref{eq:rst} gives the breakdown of the derivation for $r_{s,t}$. 

\begin{align}\label{eq:rst}
r_{s,t} & = 
\begin{cases}
1, & \text{if } t \leq 2016 \\
\frac{y_{s,2016}}{y_{s,t}}, & \text{if } t>2016 \text{ and } y_{s,2016} \text{ and } y_{s,t} \text{ are both observed} \\
\text{LinearInterp}(r_{s,\cdot}), & \text{if } t> 2016 \text{ and } y_{s,t} \text{ is missing}
\end{cases}
\end{align}
 \noindent
To ensure that estimated state-level trends reflect the observed increase in OUD rates beginning in 2016—while correcting for the artificial inflation introduced by changes in diagnostic criteria starting in 2020—we incorporate observed ratios directly into the Poisson data model. These ratios serve as lower bounds for the adjustment factors, denoted $\Delta_{s,t}$, which scale the observed case counts downward in post-2020 years. We assume the latent OUD count for each state-year falls within the range defined by the 2016 reported count and the observed values in subsequent years. Specifically, for years following the definition change ($t > t_0$), we model the adjustment factors as:

\begin{equation}
    \Delta_{s,t} \sim \text{Uniform}(r_{s,t}, 1), \quad \text{for } t > 2016
\end{equation}

\subsubsection*{Priors and Hyperpriors}
\begin{align}
    \alpha &\sim \mathcal{N}(0, 10) \nonumber\\
    \beta_j &\sim \mathcal{N}(0, 10), \quad j=1,2 \nonumber\\
    \omega_s &\sim \mathcal{N}(0, 10)\nonumber\\
    \tau_\phi &\sim \text{Gamma}(1, 100)\nonumber
\end{align}

\subsection{Post-processing of posterior estimates}

To support the second stage of our hierarchical modeling framework, we extract posterior samples of key state-level parameters from Stage I. Specifically, we obtain posterior median estimates of the latent OUD risk $\pi_{s,t}$ and the state-level random intercept $\omega_s$, which are used as inputs to the county-level model. The posterior estimates of $\omega_s$, denoted $\tilde{\omega}_s$, are used as prior information for the state-specific intercepts in Stage II. Similarly, posterior estimates of risk $\tilde{\pi}_{s,t}$ are scaled by the population size $n_{s,t}$ to compute draws of state-level OUD counts as $\tilde{y}_{s,t} = \tilde{\pi}_{s,t} \cdot n_{s,t}$, which serve as the total counts to be disaggregated across counties.
\bigskip

\noindent
\paragraph{Full Joint Posterior Distribution.} Let $\boldsymbol{y} = \{y_i\}_{i=1}^N$ denote the observed counts of opioid use disorder (OUD) for individuals in state-year pairs $(s[i], t[i])$, and let $\boldsymbol{n} = \{n_i\}_{i=1}^N$ denote the corresponding population denominators. Our model includes the following latent variables and parameters:
\[
\boldsymbol{\Theta} = \left\{\alpha, \beta_1, \beta_2, \omega_s, \phi_{s,t}, \lambda_{s,t}, \Delta_{s,t} \right\}.
\]
\noindent
The full conditional posterior distribution is proportional to the product of the likelihood and priors is given in Eq. \ref{eq:post}. This full posterior is approximated via Markov chain Monte Carlo (MCMC), from which we draw samples of all parameters in $\boldsymbol{\Theta}$.
\begin{align}\label{eq:post}
p(\boldsymbol{\Theta} \mid \boldsymbol{y}, \boldsymbol{n}) &\propto 
\prod_{i=1}^N \text{Poisson}\left(y_i \,\middle|\, \mu_i = \frac{n_i \cdot \exp(\lambda_{s[i],t[i]})}{\Delta_{s[i],t[i]}} \right)  \\
&\quad \times \prod_{s,t} \mathcal{N}\left(\lambda_{s,t} \,\middle|\, \alpha + \beta_1 x_{1,s,t} + \beta_2 x_{2,s,t} + \omega_s + \phi_{s,t}, \sigma^2_{\lambda} \right) \nonumber \\
&\quad \times \prod_s \mathcal{N}(\omega_s \mid 0, \tau_\omega^{-1}) \times \prod_{s,t} p(\phi_{s,t} \mid \phi_{s,t\pm1}, \tau_\phi) \nonumber \\
&\quad \times \prod_{s,t > t_0} \text{Uniform}(\Delta_{s,t} \mid r_{s,t}, 1) \times p(\alpha) \prod_j p(\beta_j) \cdot p(\tau_\phi).\nonumber
\end{align}

\paragraph{Posterior Estimates of \texorpdfstring{$\tilde{\pi}_{s,t}$}{pi\_st}.}

At each MCMC iteration $m = 1, \dots, M$, we draw a sample $\hat{\pi}_{s,t}^{(m)}$ from its full conditional distribution. We then compute:
\[
\tilde{\pi}_{s,t} = \text{Median} \left\{ \exp\left(\hat{\pi}_{s,t}^{(m)}\right) \right\}_{m=1}^M
\]
These posterior estimates $\tilde{\pi}_{s,t}$ are then used to generate posterior median estimates of the total number of OUD cases in state $s$ and year $t$ as:
\[
\tilde{y}_{s,t} = \tilde{\pi}_{s,t} \cdot n_{s,t}.
\]

\paragraph{Posterior Estimates of \texorpdfstring{$\tilde{\omega}_s$}{omega\_s}.}

The state-level random effects $\omega_s$ capture persistent spatial heterogeneity in OUD risk not explained by observed covariates or temporal trends. These effects are sampled within the MCMC algorithm from their full conditional distributions. At each MCMC iteration $m = 1, \dots, M$, we obtain a draw $\hat{\omega}_s^{(m)}$ from the posterior distribution of $\omega_s$, yielding a posterior sample:
\[
\left\{\hat{\omega}_s^{(1)}, \hat{\omega}_s^{(2)}, \dots, \hat{\omega}_s^{(M)}\right\}.
\]
We summarize this posterior distribution using its median, denoted $\tilde{\omega}_s = \text{Median}\left\{ \hat{\omega}_s^{(m)} \right\}_{m=1}^M$. These posterior median estimates are subsequently used as prior information for state-specific intercepts in the second-stage county-level model, where they anchor the small-area estimation procedure to the larger-scale spatial heterogeneity patterns inferred in Stage I. State Model Posterior Inference and Full Conditional Distributions are provided in Appendix~\ref{sec:posts}.

\subsection{Stage 2: County-Level Bayesian Spatial-Temporal Model}
In Stage II, the county-level counts of OUD cases, denoted $\mu_{c,t}$ for county $c$ in state $s[c]$ and year $t$, are unobserved. To enable inference at this finer spatial resolution, we adopt a foundational data-generating assumption: for each state $s$ and year $t$, the sum of the latent county-level counts within state $s$ must equal the estimated state-level total $\tilde{y}_{s,t}$. This constraint ensures internal consistency across spatial scales and preserves coherence of OUD burden between county and state resolutions. Accordingly, we assume that the latent counts across counties within each state follow a multinomial distribution:
\begin{equation}\label{eq:multi}
\left\{\mu_{c,t} : c \in C_s\right\} \sim \text{Multinomial}\left(\left\{\rho_{c,t} : c \in C_s\right\}, \tilde{y}_{s,t} \right),
\end{equation}
where $\rho_{c,t}$ denotes the probability that a case in state $s$ and year $t$ is allocated to county $c$, and $\sum_{c \in C_s} \rho_{c,t} = 1$. This modeling assumption enforces the constraint:
\begin{equation*}
\sum_{c \in C_s} \mu_{c,t} = \tilde{y}_{s,t},
\end{equation*}
ensuring that county-level latent counts sum exactly to the estimated state-level total for each $s$ and $t$.

\paragraph{Data Model}
To facilitate modeling and computation, particularly when incorporating covariates and spatial structure, we adopt an equivalent formulation of this multinomial model using a set of Poisson distributions. Specifically, we model each county-year count as:

\begin{align}
\mu_{c,t} & \sim \text{Poisson}(\mu_{c,t}), \quad \text{where } \mu_{c,t} = \rho_{c,t} \cdot \tilde{y}_{s,t},\\
\rho_{c,t}&  = \frac{\exp(\eta_{c,t})}{\sum_{c' \in C_s} \exp(\eta_{c',t})},\nonumber
\end{align}

so that $\rho_{c,t}$ represents the softmax-normalized allocation of $\tilde{y}_{s,t}$ across counties \citep{baker1994multinomial, baker2008generalized}. Under this formulation, the Poisson observations are independent, and the sum of expected counts across counties satisfies:

\[
\sum_{c \in S} \mu_{c,t} = \sum_{c \in S} \rho_{c,t} \cdot \tilde{y}_{s,t} = \tilde{y}_{s,t}.
\]

While the Poisson and multinomial models are not equivalent marginally, they are conditionally equivalent: if $\mu_{c,t} \sim \text{Poisson}(\mu_{c,t})$ independently, then the distribution of $\{\mu_{c,t}\}$ conditional on $\sum_{c} \mu_{c,t} = \tilde{y}_{s,t}$ follows a multinomial distribution with total $\tilde{y}_{s,t}$ and probabilities $\rho_{c,t}$. This reparameterization preserves the foundational property that county-level counts must sum to the state-level total, while enabling more flexible specification of $\eta_{c,t}$ using generalized linear modeling frameworks (e.g., log-linear models with spatial random effects). It also improves computational tractability in Bayesian frameworks by allowing for conditionally independent likelihood contributions at the county level.

\paragraph*{Population-weighted correction.}
To improve stability in the allocation of state-level totals $\tilde{y}_{s,t}$ across counties, we apply a population-weighted correction to the softmax normalization used in computing $\rho_{c,t} = \mu_{c,t}/\tilde{y}_{s,t}$. This correction is necessary when county population sizes vary widely, as small populations can otherwise lead to extreme or unstable allocation probabilities. We define a log-scale offset for each county-year pair as:
\[
\text{log\_offset}_{c,t} = \log(\tilde{y}_{s,t}) + \log(N_{c,t}) - \log\left(\sum_{c' \in C_s} N_{c',t}\right),
\]
where $N_{c,t}$ is the population-at-risk (population age 12+) for county $c$ and year $t$, and $\tilde{y}_{s,t}$ is the posterior mean of the total state-level OUD count. This log-scale correction approximates a population-weighted allocation of the total state-level burden $\tilde{y}_{s,t}$ under the assumption that, in the absence of covariates and spatial effects, the expected burden in each county should be proportional to its population. That is, if we assume $\mu_{c,t} \propto N_{c,t}$, then the expected share of the total burden becomes:
\[
\rho_{c,t} = \frac{N_{c,t}}{\sum_{c' \in C_s} N_{c',t}}.
\]
Taking the log of this expression gives:
\[
\log(\rho_{c,t}) = \log(N_{c,t}) - \log\left(\sum_{c' \in C_s} N_{c',t} \right).
\]
Because $\rho_{c,t}$ is defined through a softmax of an unnormalized log-linear predictor, we can incorporate this term as an additive offset. To preserve scaling to the total count $\tilde{y}_{s,t}$, we add $\log(\tilde{y}_{s,t})$, which aligns the expected counts $\mu_{c,t}$ with the state-level total:
\[
\log(\mu_{c,t}) = \log(\tilde{y}_{s,t}) + \log(N_{c,t}) - \log\left(\sum_{c' \in C_s} N_{c',t} \right).
\]
Exponentiating both sides gives $\mu_{c,t} = \tilde{y}_{s,t} \cdot \frac{N_{c,t}}{\sum_{c' \in C_s} N_{c',t}}$ which is the desired population-weighted allocation. Thus, the proposed log-scale offset ensures that, in the absence of additional model structure, the Poisson means $\mu_{c,t}$ sum to the state-level total and are proportionally distributed by population size.

\paragraph{Latent Process Model.}

The county-level log-risk $\eta_{c,t}$ characterizes the relative burden of OUD in county $c$ and year $t$, and is modeled using a log-linear specification:

\begin{align}
\eta_{c,t} & = \alpha + \sum_{j=1}^J \beta_j x_{j,c,t} + \phi_{c,t} + \tilde{\omega}_s,\\
\gamma_s & \sim \mathcal{N}(\tilde{\omega}_s, \tilde{\sigma}_{\omega}^2)\nonumber\\
\alpha, \beta_j & \sim \mathcal{N}(0, 1)\nonumber
\end{align}

where $x_{j,c,t}$ denotes the $j^{\text{th}}$ covariate for county $c$ in year $t$, capturing structural risk factors including: mortalty from any opioid use, rurality, poverty, disability, and the cumulative opioid prescription rate. $\beta_j$ are fixed-effects coefficients corresponding to each covariate, and $\alpha$ is the global intercept. The state-level random effect $\gamma_s$ is normally distributed centered around the posterior estimate of the state-level intercept $\tilde{\omega}_s$ and corresponding variance $\tilde{\sigma}_{\omega_s}$ obtained from Stage I, incorporated as an offset term to propagate state-level information into the county-level model. This latent process governs the relative probability $\rho_{c,t}$ used in the Poisson likelihood. The inclusion of $\tilde{\omega}_s$ ensures coherence between the two model stages by anchoring county-level predictions to the corresponding state-level estimates. By integrating over the posterior distributions of $\tilde{y}_{s,t}$ and $\tilde{\omega}_s$, this formulation yields fully probabilistic, internally consistent estimates of county-level OUD burden that respect the hierarchical structure of the data. 
\bigskip

\paragraph{Spatio-Temporal Random Effect Specification.}

To account for both spatial correlation across counties and temporal evolution in unmeasured risk factors, we model the latent spatio-temporal random effect $\phi_{c,t}$ as the sum of a spatially structured effect and a temporally structured deviation:
\[
\phi_{c,t} = u_c + \delta_{c,t},
\]
where $u_c$ captures spatial dependence across counties and $\delta_{c,t}$ accounts for residual temporal variation within each county.

The spatial effect $u_c$ is modeled using a Besag-York-Mollié (BYM) specification, which includes both spatially structured and unstructured components:
\[
u_c = u_c^{\text{str}} + u_c^{\text{unstr}},
\]
with:
\begin{align*}
u_c^{\text{str}} \mid \{u_{c'}^{\text{str}}\}, \tau_u &\sim \mathcal{N}\left( \frac{1}{N_c} \sum_{c' \sim c} u_{c'}^{\text{str}}, \frac{1}{\tau_u N_c} \right), \\
u_c^{\text{unstr}} &\sim \mathcal{N}(0, \tau_v^{-1}),
\end{align*}
where $c' \sim c$ denotes that counties $c$ and $c'$ are neighbors, $N_c$ is the number of neighbors of county $c$, and $\tau_u$, $\tau_v$ are precision parameters for the structured and unstructured components, respectively.

The temporal deviation $\delta_{c,t}$ is modeled as a first-order random walk (RW1) process within each county:
\[
\delta_{c,t} \mid \delta_{c,t-1}, \tau_{\delta} \sim \mathcal{N}(\delta_{c,t-1}, \tau_{\delta}^{-1}),
\]
where $\tau_\delta$ controls the smoothness of temporal evolution.

Together, this additive formulation allows the model to capture long-term spatial heterogeneity across counties as well as gradual temporal changes in local OUD risk that are not explained by covariates. The structure supports borrowing strength both across neighboring counties and across adjacent time points, yielding robust county-year estimates in the presence of data sparsity. 

\paragraph*{County-level risk as the target quantity.}
The primary quantity of interest in this analysis is the county-level OUD risk, denoted $\pi_{c,t}$, defined as the proportion of individuals in county $c$ and year $t$ affected by OUD. Although the model is formulated using a Poisson likelihood with allocation proportions $\rho_{c,t}$, these proportions serve as intermediate quantities used to disaggregate the estimated state-level total $\tilde{y}_{s,t}$ into county-level counts. We then compute the estimated county-level risk by dividing the expected count $\mu_{c,t} = \rho_{c,t}\cdot \tilde{y}_{s,t}$ by the known county population $N_{c,t}$, i.e., $\pi_{c,t} = \frac{\mu_{c,t}}{N_{c,t}}$.
This formulation allows us to generate interpretable, spatially detailed, and fully probabilistic estimates of OUD burden at the county-year level. 

\subsection{Simulation Design}\label{sec:simulation}
\paragraph*{Data Generation}
We generated $M=100$ simulated datasets using the following procedure. First, we fixed global parameters from the county-level process model (e.g., $\alpha$, $\beta_j$, spatial and temporal variance components) and used real observed covariates $x_{j,c,t}$ from the application dataset. For each simulation, we generated spatio-temporal random effects $\phi_{c,t}$ from their respective prior distributions (e.g., ICAR for spatial effects, random walk for temporal trends). We then computed the true latent log-risk for each county-year and derived the true incidence risk via the inverse log-link. County-level counts were generated using the Poisson likelihood where $n_{c,t}$ is the known population denominator. Lastly, state-level counts were computed by summing across counties. The summary of the data generating mechanism is shown in Eq. \ref{eq:sim}.
\begin{align}\label{eq:sim}
\eta_{c,t,m}^{\text{true}} &= \alpha^{\text{fixed}} + \sum_{j=1}^J \beta_j^{\text{fixed}}  x_{j,c,t} + \phi^{\text{true}}_{c,t,m},\\
\pi_{c,t,m}^{\text{true}} &= \exp(\eta_{c,t,m}^{\text{true}})\nonumber\\
\mu_{c,t,m}^{\text{sim}}& \sim \text{Poisson}\left( \pi_{c,t,m}^{\text{true}} \cdot n_{c,t} \right),\nonumber\\
y_{s,t,m}^{\text{sim}} &= \sum_{c \in S} \mu_{c,t,m}^{\text{sim}}\nonumber
\end{align}

Each of the 100 simulated datasets consisted of synthetic county- and state-level counts $\{\mu_{c,t,m}^{\text{sim}}, y_{s,t,m}^{\text{sim}}\}$, paired with the observed covariates and known population sizes.

\paragraph*{Model Fitting and Evaluation}

For each simulated dataset $m = 1, ..., M$, we applied the full two-stage modeling pipeline:
\begin{enumerate}
    \item Fit the Stage I level of the B-Step approach to generated $y_{s,t,m}^{sim}$ to estimate $\tilde{\pi}_{s,t}^{(m)}$ and $\tilde{y}_{s,t}^{(m)}$,
    \item Use the posterior estimates $\tilde{y}_{s,t}^{(m)}$ and $\tilde{\omega}_s^{(m)}$ in Stage II to estimate $\tilde{\pi}_{c,t}^{(m)}$.
\end{enumerate}

We compared the posterior median estimates $\tilde{\pi}_{s,t}$ and $\tilde{\pi}_{c,t}$ to the true simulated risks $\pi_{s,t}^{\text{true}}$ and $\pi_{c,t}^{\text{true}}$. For each parameter at both spatial scales (state and county), we computed the following performance metrics across all simulation replicates:

These metrics were calculated separately for each simulation iteration and summarized across all counties and states.

\begin{mdframed}[backgroundcolor=gray!10, linewidth=0.5pt, roundcorner=5pt]
\textbf{Simulation Summary:}
\begin{itemize}
    \item \textbf{Step 1:} Fix global model parameters from the county-level model.
    \item \textbf{Step 2:} Use observed covariates and simulated spatial (ICAR) and temporal (RW1) effects to construct $\eta_{c,t}^{\text{true}}$ and $\pi_{c,t}^{\text{true}}$.
    \item \textbf{Step 3:} Generate synthetic counts $\mu_{c,t}^{\text{sim}} \sim \text{Poisson}(\pi_{c,t}^{\text{true}} \cdot n_{c,t})$.
    \item \textbf{Step 4:} Aggregate to state-level counts $y_{s,t}^{\text{sim}} = \sum_c \mu_{c,t}^{\text{sim}}$.
    \item \textbf{Step 5:} Fit the two-stage model to each dataset to estimate $\tilde{\pi}_{s,t}$ and $\tilde{\pi}_{c,t}$.
    \item \textbf{Step 6:} Compare estimated quantities to $\pi_{s,t}^{\text{true}}$ and $\pi_{c,t}^{\text{true}}$ using ME, MdE, MAE, MRE, MSE, and 95\% coverage.
    \begin{itemize}
    \item \textbf{Mean Error (ME)}: $\frac{1}{M} \sum_{m=1}^M \left( \tilde{\pi}^{(m)} - \pi^{\text{true}} \right)$
    \item \textbf{Median Error (MdE)}: median of $\tilde{\pi}^{(m)} - \pi^{\text{true}}$
    \item \textbf{Median Absolute Error (MAE)}: median of $|\tilde{\pi}^{(m)} - \pi^{\text{true}}|$
    \item \textbf{Median Relative Error (MRE)}: median of $\left| \frac{\tilde{\pi}^{(m)} - \pi^{\text{true}}}{\pi^{\text{true}}} \right|$
    \item \textbf{Mean Squared Error (MSE)}: $\frac{1}{M} \sum_{m=1}^M \left( \tilde{\pi}^{(r)} - \pi^{\text{true}} \right)^2$
    \item \textbf{Coverage Probability}: proportion of simulations for which the 95\% posterior credible interval for $\pi$ contains $\pi^{\text{true}}$
\end{itemize}
\end{itemize}
\end{mdframed}

\section{Computation}
We obtained population denominators and associated uncertainty estimates from the American Community Survey (ACS) and Population Estimates Program (PEP) using the tidycensus R package \citep{walker_k_tidycensus_2020}. Stage I model estimation was conducted using Markov Chain Monte Carlo (MCMC) sampling via the NIMBLE software framework \citep{perry_de_valpine_programming_nodate}, enabling full posterior inference for all model parameters. We ran eight parallel chains with 80,000 iterations each, discarding the first 40,000 iterations as burn-in. Convergence diagnostics—including visual inspection of traceplots and rank-based metrics—were used to ensure adequate mixing and convergence of posterior samples \citep{a_gelman_inference_1992, gabry_visualization_2019, hastie_model_2009,  plummer_jags_2017,   aki_vehtari_rank-normalization_2021}. In Stage II, posterior estimates from Stage I were incorporated into a spatial disaggregation model implemented via integrated nested Laplace approximation (INLA) using the R-INLA package \citep{rue2009inla, r-inla, havard_rue_gaussian_2005}.

\section{Results}\label{sec:results}
\subsection{State-Level Results}
\subsubsection{Global Parameter Estimates}
The posterior summaries of global parameters from Stage I of B-Step are presented in Table~\ref{tab:posterior_summary_stage1}. The global intercept $\alpha$ had a posterior mean of $-4.69$ (95\% CrIs: $[-4.87, -4.38]$), indicating a low baseline log incidence risk of OUD across all states and years. The coefficient for PR misuse ($\beta_1$) was negative on average ($\text{mean} = -1.58$), but showed high uncertainty with a wide credible interval spanning both negative and positive values ($[-6.07, 1.45]$), suggesting limited precision in estimating its effect. In contrast, the coefficient for heroin use ($\beta_2$) had a positive mean ($0.79$) with a moderately wide credible interval ($[-1.17, 2.77]$), indicating potential association with higher OUD incidence. The posterior standard deviation of the state-specific temporal random walk component ($\sigma_{\text{RW}}$) was relatively small ($0.079$), suggesting modest temporal deviation around the modeled trend.

\begin{table}[ht]
\centering
\caption{Posterior summaries of global model parameters from B-Step Stage I.}
\begin{tabular}{|l|r|r|r|r|r|}
\hline
\textbf{Parameter} & \textbf{Mean} & \textbf{Median} & \textbf{Std. Dev.} & \textbf{95\% CI Low} & \textbf{95\% CI Upp} \\
\hline
$\alpha$    & -4.6880 & -4.7190 & 0.1340 & -4.8715 & -4.3789 \\
$\beta_1$   & -1.5828 & -1.2551 & 2.0174 & -6.0747 &  1.4518 \\
$\beta_2$   &  0.7949 &  0.7928 & 1.0051 & -1.1650 &  2.7715 \\
$\sigma_{\text{RW}}$ &  0.0786 &  0.0672 & 0.0585 &  0.0029 &  0.2183 \\
\hline
\end{tabular}
\label{tab:posterior_summary_stage1}
\end{table}

\subsubsection{Posterior summaries of state-level trends in OUD risk}
\begin{figure}[H]
  \centering
  \includegraphics[width=0.6\textwidth]{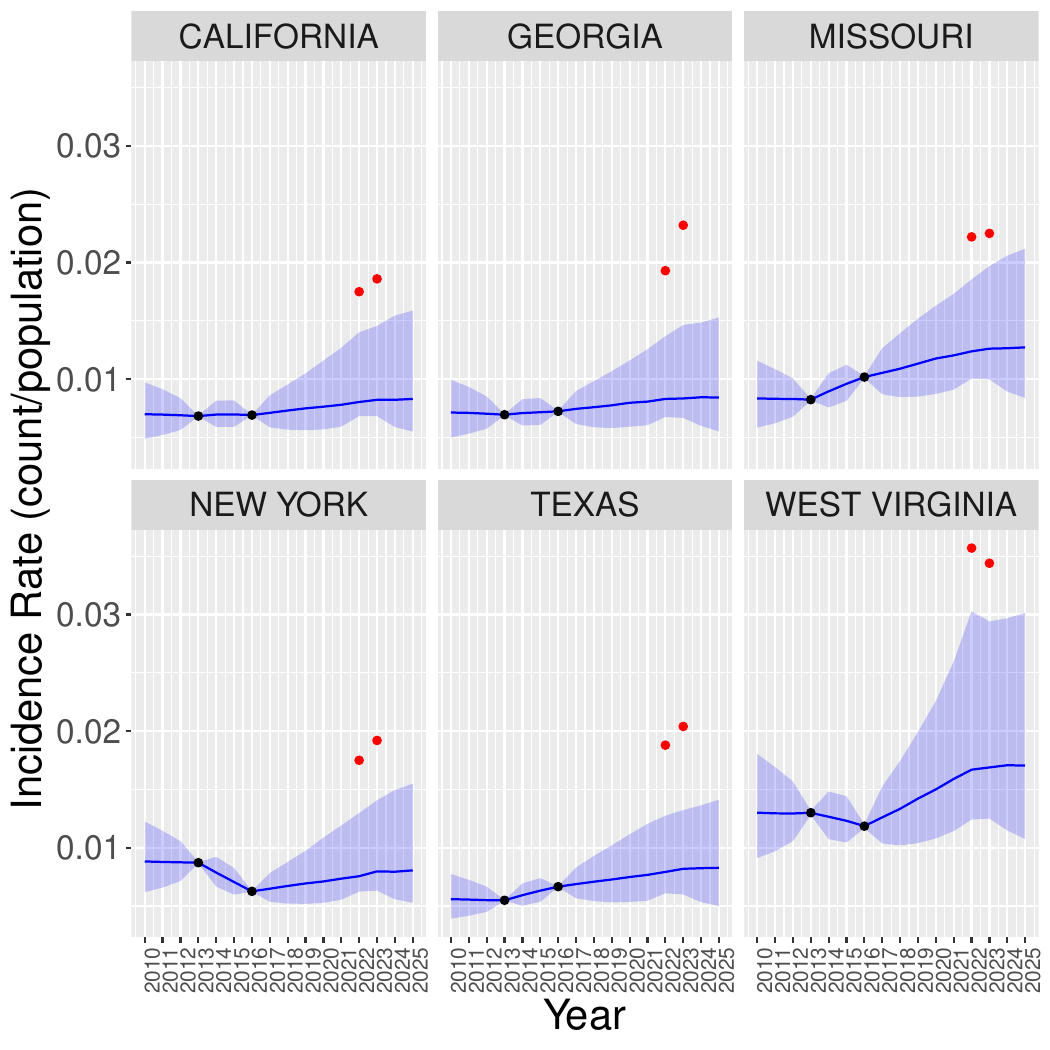}
  \caption{State-level trends in posterior median risk estimates and 95\% CrIs for selected states. Blue lines and shaded areas denotes median estimates and uncertainty bounds. The black dots denote data prior to 2020, and red dots post 2020.}
  \label{fig:statetrends}
\end{figure}

Figure~\ref{fig:statetrends} presents estimated risk trends for selected states from 2010 to 2025. States with historically high opioid-related mortality, such as West Virginia, maintained consistently elevated risk throughout the study period, while others experienced more gradual increases. Beginning in 2016, a definitional adjustment was applied to account for changes in case identification, resulting in estimated risk that rise gradually but remain lower than observed values. Posterior uncertainty, represented by the width of the 95\% credible intervals (95\% CrIs), was narrowest around 2013 and 2016, and increased substantially in the years following 2016. This reflects both the added uncertainty introduced by the definition adjustment and the growing temporal distance from 2016—the last year with fully observed data.

\subsubsection{Estimates of Adjustment Factors}
\begin{figure}[htb]
  \centering
  \includegraphics[width=0.4\textwidth]{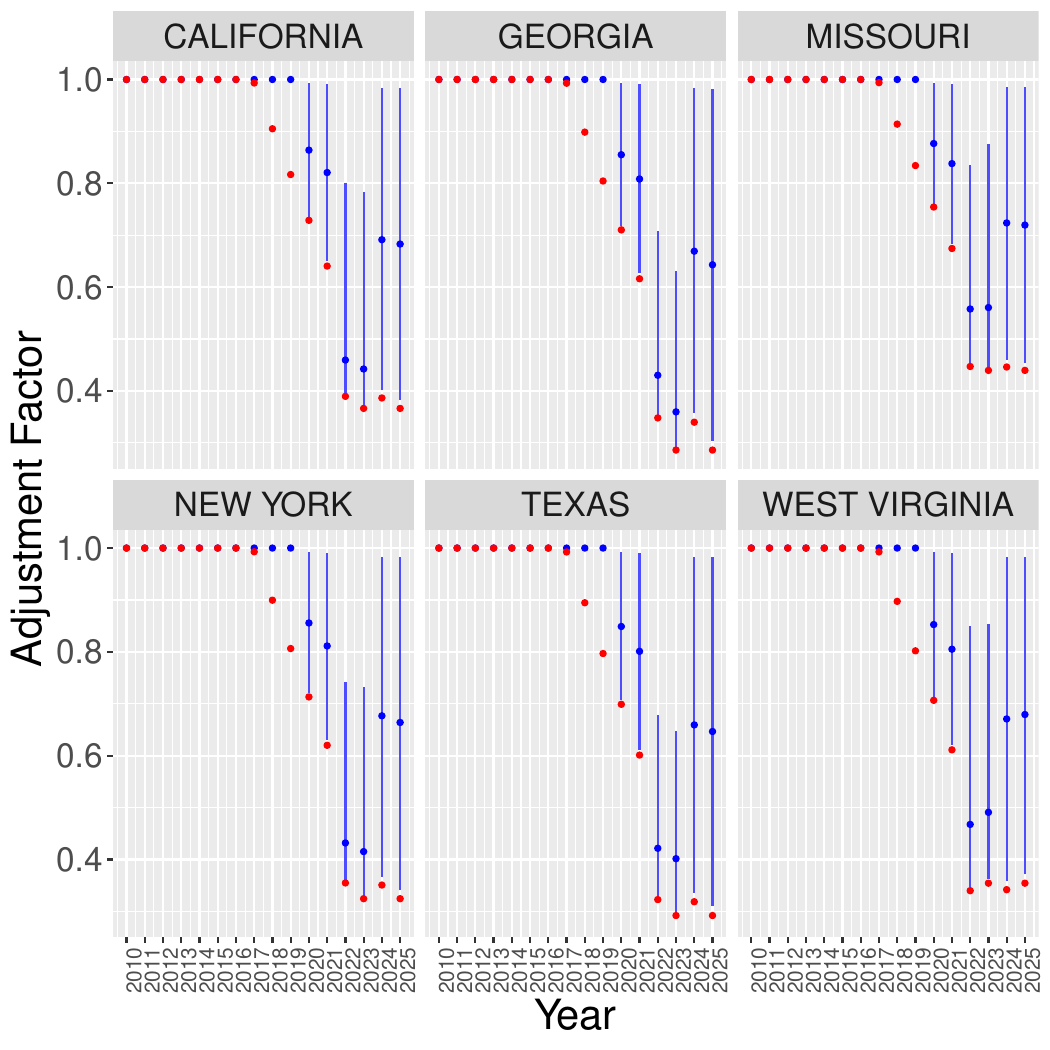}
  \caption{State-level trends in adjustment factors. Red dots denote the reported $r_{s,t}$, i.e., lower bound of the Uniform distribution. Blue dots denote median adjustment estimates, and blue lines denote 95\% CrIs.}
  \label{fig:adj}
\end{figure}

Figure~\ref{fig:adj} displays the estimated adjustment factors $\Delta_{s,t}$ over time for selected states (blue). Each factor is assigned a prior distribution $\Delta_{s,t} \sim \text{Uniform}(r_{s,t}, 1)$, where the lower bound $r_{s,t}$ represents the observed ratio of OUD incidence risk in year $t$ relative to 2016, denoted with the red dots. For years $t \leq 2016$, $\Delta_{s,t}$ is fixed at 1, indicating no adjustment is necessary.  From 2017 through 2025, the estimated adjustment factors generally decline, capturing inflation in observed case counts due to expanded diagnostic criteria. In all states, the estimated adjustment $\tilde{\Delta}_{s,t}$ for years 2022-2023 are close to the lower bound of the uniform distribution, i.e., close to the observed ratio. For years beyond 2023, where no surveillance data are available, $\Delta_{s,t}$ is inferred as the midpoint of its prior interval, effectively reflecting prior uncertainty. As a result, post-2023 incidence trends are driven by the temporal random walk component, which extrapolates existing patterns in a data-consistent manner.

\subsection{County Level Results}
\subsubsection{Posterior summaries of county-level trends in OUD risk}
\begin{figure}[H]
  \centering
  \includegraphics[width=0.85\textwidth]{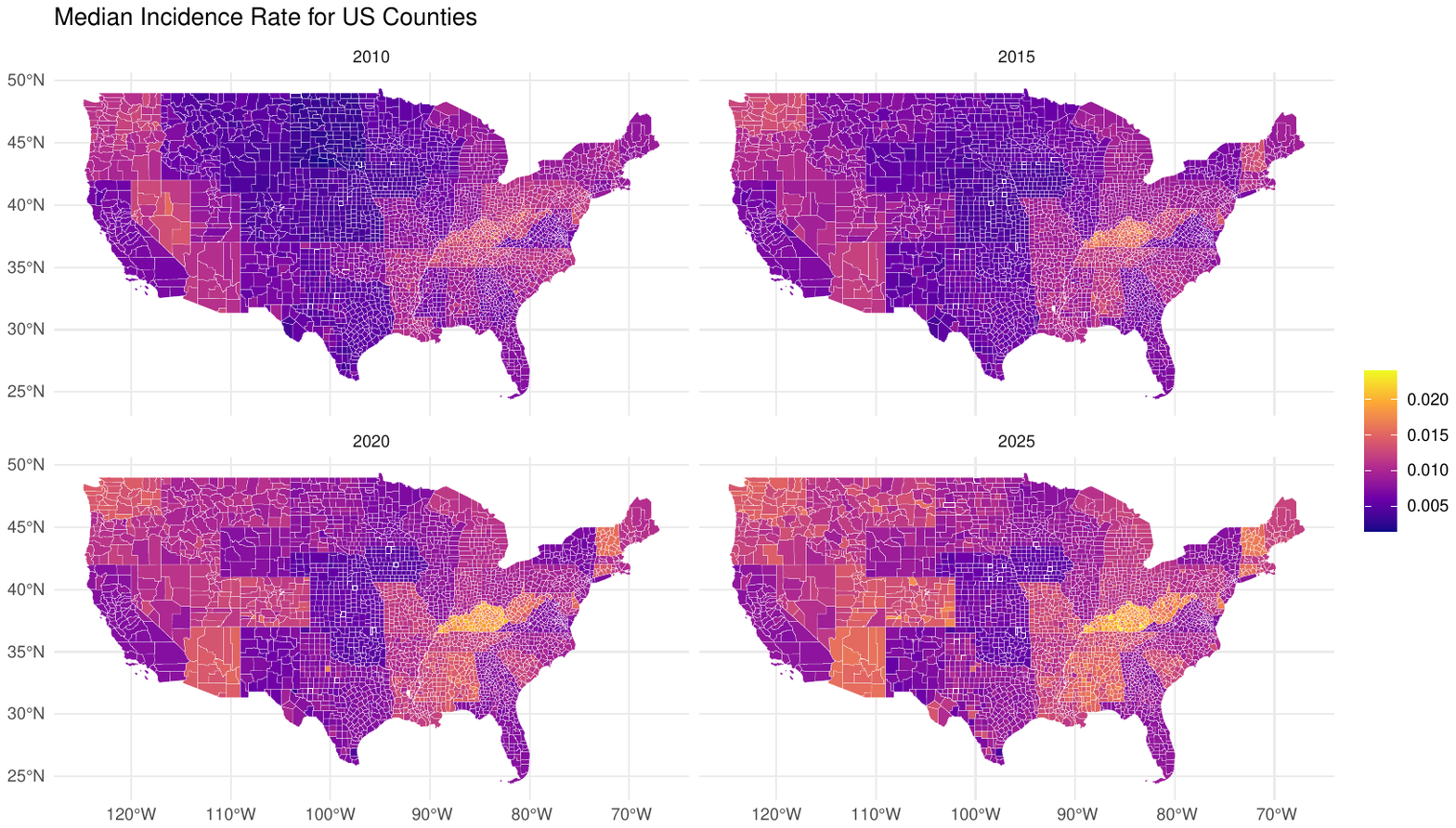}
  \caption{Mapped county-level posterior median risk estimates $\tilde{\pi}_{c,t}$ for years 2010, 2015, 2020, 2025. Blue denotes lower risk vs. yellow of higher risk probabilities. Counties were subsetted to include on mainland U.S. counties for ease of readability. The full county level estimates can be found in Appendix \ref{sec:countyfull}. }
  \label{fig:countymap}
\end{figure}

Figure~\ref{fig:countymap} maps the county-level posterior median  estimates of risk across all U.S. counties for the years 2010, 2015, 2020, and 2025. The figure illustrates substantial geographic variation in county-level OUD risk nationwide. In 2010, the lowest estimated risk occurred in Mellette County, South Dakota (population 1,585), with a median rate of 0.0015 [95\% CrIs: (0.0014, 0.0016)]. In contrast, the highest risk occurred in Doddridge County, West Virginia (population 7,030), with a median risk of 0.018 [95\% CrIs: (0.017, 0.019)]. By 2020, the lowest risk was in Loving County, Texas, with a median risk of 0.0026 [95\% CrIs: (0.0021, 0.0033)], while the highest was in Robertson County, Kentucky, at 0.024 [95\% CrIs: (0.022, 0.026)]. The persistence of both high- and low-risk areas over time underscores ongoing disparities in opioid burden and the role of localized social and structural determinants. These extremes help identify regions with the greatest need for targeted intervention and continued surveillance.

\subsubsection{Temporal Trends in County-Level Uncertainty}
\begin{figure}[htb]
  \centering
  \includegraphics[width=0.9\textwidth]{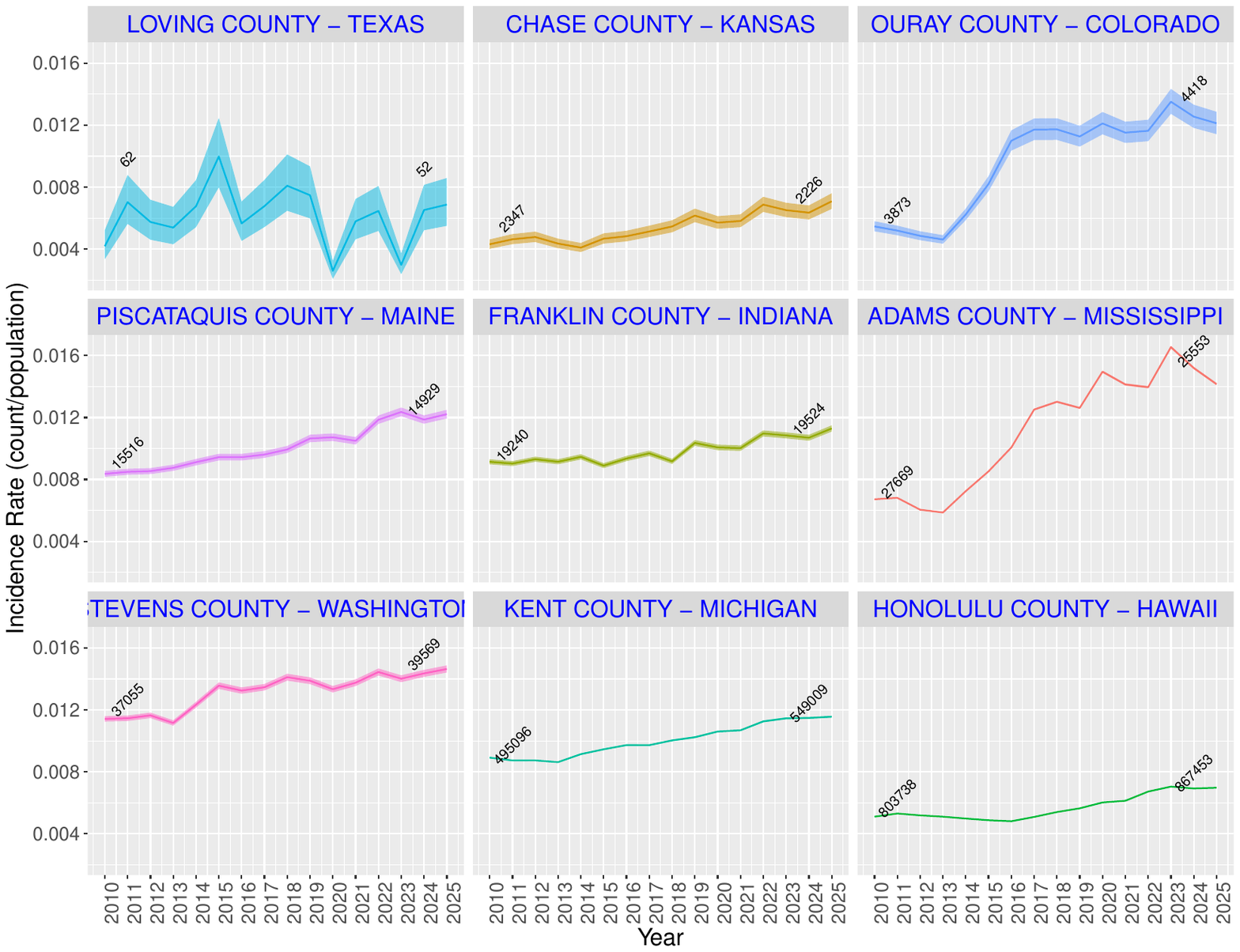}
  \caption{Temporal trends in county-level posterior median OUD risk estimates with 95\% credible intervals for selected counties illustrating results for varying population sizes. Numeric calues at the beginning and end of the temporal trend denote US Census reported population sizes in 2010 and 2023.}
  \label{fig:countymir}
\end{figure}

Figure~\ref{fig:countymir} presents temporal trends in county-level OUD risk estimates alongside their associated 95\% CrIs for selected counties, highlighting how uncertainty varies with population size. Loving County, Texas—one of the least populous counties in the U.S.—had a population of just 62 in 2010 and 52 in 2023. As expected, the corresponding 95\% CrIs are wide, reflecting substantial uncertainty in the estimated OUD risk due to sparse data. A moderately sized county, e.g., Piscataquis County, Maine, with a population size of 14,929 in 2023, shows a greatly reduced degree of uncertainty compared to the previous small county example. Honolulu County, Hawaii, with a population exceeding 800,000 across the same period, exhibits very narrow CrIs, indicating greater precision in the estimated risk. These comparisons underscore the influence of population size on the uncertainty of small-area estimates and highlight the importance of fully probabilistic modeling approaches in settings with heterogeneous data density.

For comprehensive county level risk estimates refer to Appendix Sections \ref{sec:countyfull}- \ref{sec:discs}, which show full graphical representations of county estimates within each state. 

\subsection{Simulation Results}
To evaluate the performance of the proposed B-Step model, we applied the full two-stage modeling pipeline to each of the $M=100$ simulated datasets as described in Section~\ref{sec:simulation}. We assessed model performance by comparing the posterior median estimates of OUD risk to the known ground truth at both the state and county levels. Table~\ref{tab:sim_results} summarizes the performance metrics computed across all simulations for state-level and county-level OUD risk estimates, including mean error (ME), median error (MdE), median absolute error (MAE), median relative error (MRE), mean squared error (MSE), and 95\% coverage probability. Results are presented as averages across simulations and spatial units.

\begin{table}[H] \centering \caption{Simulation performance metrics comparing posterior estimates $\tilde{\pi}$ to true risks $\pi^{\text{true}}$ across 100 simulated datasets.} 
\label{tab:sim_results} 
\begin{tabular}{lcccccc} 
\toprule 
\textbf{Scale} & \textbf{ME} & \textbf{MdE} & \textbf{MAE} & \textbf{MRE} & \textbf{MSE} & \textbf{Coverage} \\
\hline
State-level & 0.0008 & 0.0005 & 0.0031 & 0.052 & 0.013& 0.93 \\
\hline
County-level & 0.004 & 0.002 & 0.049 & 0.84 & 0.024 & 0.89 \\
\hline
\bottomrule 
\end{tabular} 
\end{table}

At the state-level, the B-Step model achieved high accuracy, with negligible bias (ME = 0.0008), tight median absolute error (MAE = 0.0031), and close to expected 93\% coverage as expected. The slightly higher median relative error (MRE) at the county level reflects greater uncertainty due to smaller population sizes and the absence of directly observed data, but the model still maintained robust performance and high coverage. To further examine the behavior of the B-Step model, we evaluated how estimation uncertainty varied by county population. Figure~\ref{fig:countymir} displays posterior median estimates and 95\% CrIs for selected counties, illustrating the expected relationship: counties with larger populations (e.g., Honolulu County, HI) showed narrower CrIs, while those with very small populations (e.g., Loving County, TX) exhibited substantially wider intervals. This pattern is consistent with the model's hierarchical design, which borrows strength across space and time to stabilize estimates in data-sparse settings, while appropriately reflecting uncertainty where information is limited. Finally, we assessed the calibration of posterior uncertainty by computing empirical coverage probabilities. As shown in Table~\ref{tab:sim_results}, the 95\% credible intervals achieved near-nominal coverage at both the state and county levels, indicating well-calibrated uncertainty quantification. This supports the appropriateness of the two-stage Bayesian approach in propagating and representing uncertainty under data sparsity.

\section{Discussion}\label{sec:discussion}

In this study, we developed and validated a two-stage Bayesian spatio-temporal top-down modeling framework (B-Step) to generate small-area estimates of opioid use disorder (OUD) risk across 3,143 U.S. counties from 2010 to 2025. This approach addresses critical challenges in substance use surveillance, particularly when direct small-area data are unavailable, unreliable, or subject to definitional inconsistencies over time. Our model introduces methodological innovations that enhance the stability, interpretability, and scalability of small-area risk estimation under data sparsity. First, our results demonstrate the utility of a hierarchical top-down disaggregation strategy that infers county-level outcomes from state-level surveillance data using a combination of population scaling, covariate information, and spatial-temporal random effects. By adopting a softmax-normalized Poisson formulation, we ensure that county-level estimates coherently aggregate to state-level totals while allowing flexible incorporation of local covariates and random effects. The model retains full probabilistic structure across both levels, enabling us to quantify uncertainty and propagate it through the estimation pipeline. Second, we incorporated a population-weighted offset term into the softmax specification to mitigate instability in scaling factors, particularly in sparsely populated counties. This log-scale correction ensures that, in the absence of covariates or spatial structure, expected county-level burdens are proportionally allocated according to population size. As shown in simulation and empirical results, this adjustment reduces spurious variability and improves the interpretability of small-area estimates. Third, our model directly incorporates an adjustment factor to address inconsistencies in OUD case definitions introduced in 2020, when the NSDUH expanded its diagnostic criteria. By placing a uniform prior over an empirically derived ratio of pre- and post-2020 case counts, we ensure that post-2020 trends reflect true temporal evolution in OUD burden rather than artificial increases due to definitional changes. This adjustment mechanism is novel in the context of OUD modeling and provides a generalizable strategy for harmonizing surveillance data in the presence of shifting diagnostic or reporting practices. Lastly, we adopted a modular Bayesian approach, ensuring that the Stage I model for state-level estimation is informed solely by state-level data and priors, without feedback from the county-level model. This preserves identifiability and prevents circular information flow. Modular inference has been increasingly recommended in hierarchical systems to avoid overfitting and reduce computational complexity, and our implementation shows its practical benefits in public health surveillance contexts. The B-Step model was evaluated through extensive simulation studies that demonstrated strong accuracy and calibration. The model achieved near-nominal 95\% coverage at both state and county levels, low bias, and stable performance across population sizes. These results validate our framework's ability to recover true latent risks even in highly sparse and heterogeneous data settings.

From a public health perspective, the model offers actionable insights for policymakers and practitioners. Our application revealed persistent spatial disparities in county-level OUD risk, with high-burden counties concentrated in Appalachia and parts of the Midwest. Moreover, the widening uncertainty bounds in low-population counties highlight the importance of investing in targeted data collection in underserved areas to reduce uncertainty and improve intervention precision. Additionally, the development of the B-Step model approach offers an adaptable model framework that can be used for further monitoring of small area OUD risks, as well as other applications that suffer from similar challenges of data sparsity and bias. Future research could explore joint modeling of additional indicators such as opioid-related mortality, treatment admissions, and polysubstance use, extending B-Step to a multivariate framework. Additionally, integration with spatial capture-recapture methods and data fusion techniques could enhance detection of hidden or marginalized subpopulations. Finally, model outputs could inform decision-support tools for real-time resource allocation, forecasting, and program evaluation. In conclusion, our proposed B-Step framework provides a principled, flexible, and interpretable methodology for estimating small-area opioid use disorder risk under conditions of data sparsity and structural surveillance challenges. As the opioid crisis continues to evolve, data-driven tools such as ours are essential to ensure equitable and efficient public health responses.

\clearpage

 \bibliography{MyLib}

\begin{thebibliography}{52}
\expandafter\ifx\csname natexlab\endcsname\relax\def\natexlab#1{#1}\fi
\providecommand{\url}[1]{\texttt{#1}}
\providecommand{\href}[2]{#2}
\providecommand{\path}[1]{#1}
\providecommand{\DOIprefix}{doi:}
\providecommand{\ArXivprefix}{arXiv:}
\providecommand{\URLprefix}{URL: }
\providecommand{\Pubmedprefix}{pmid:}
\providecommand{\doi}[1]{\href{http://dx.doi.org/#1}{\path{#1}}}
\providecommand{\Pubmed}[1]{\href{pmid:#1}{\path{#1}}}
\providecommand{\bibinfo}[2]{#2}
\ifx\xfnm\relax \def\xfnm[#1]{\unskip,\space#1}\fi
\bibitem[{{A Gelman} and {DB Rubin}(1992)}]{a_gelman_inference_1992}
\bibinfo{author}{{A Gelman}}, \bibinfo{author}{{DB Rubin}}, \bibinfo{year}{1992}.
\newblock \bibinfo{title}{Inference from iterative simulation using multiple sequences (with discussion)}.
\newblock \bibinfo{journal}{Statistical Science} \bibinfo{volume}{7}, \bibinfo{pages}{457--472}.
\bibitem[{Arambepola et~al.(2022)Arambepola, Lucas, Nandi, Gething and Cameron}]{arambepola_simulation_2022}
\bibinfo{author}{Arambepola, R.}, \bibinfo{author}{Lucas, T.C.D.}, \bibinfo{author}{Nandi, A.K.}, \bibinfo{author}{Gething, P.W.}, \bibinfo{author}{Cameron, E.}, \bibinfo{year}{2022}.
\newblock \bibinfo{title}{A simulation study of disaggregation regression for spatial disease mapping}.
\newblock \bibinfo{journal}{Statistics in Medicine} \bibinfo{volume}{41}, \bibinfo{pages}{1--16}.
\newblock \URLprefix \url{https://onlinelibrary.wiley.com/doi/10.1002/sim.9220}, \DOIprefix\doi{10.1002/sim.9220}.
\bibitem[{Baker(1994)}]{baker1994multinomial}
\bibinfo{author}{Baker, S.G.}, \bibinfo{year}{1994}.
\newblock \bibinfo{title}{The multinomial-poisson transformation}.
\newblock \bibinfo{journal}{The Statistician} \bibinfo{volume}{43}, \bibinfo{pages}{495--504}.
\newblock \URLprefix \url{https://www.jstor.org/stable/2348134}, \DOIprefix\doi{10.2307/2348134}.
\bibitem[{Baker(2008)}]{baker2008generalized}
\bibinfo{author}{Baker, S.G.}, \bibinfo{year}{2008}.
\newblock \bibinfo{title}{Generalized self-consistency: Multinomial logit model and poisson log-linear model}.
\newblock \bibinfo{journal}{Computational Statistics \& Data Analysis} \bibinfo{volume}{52}, \bibinfo{pages}{4483--4494}.
\newblock \URLprefix \url{https://www.sciencedirect.com/science/article/pii/S0378375807003734}, \DOIprefix\doi{10.1016/j.csda.2007.02.009}.
\bibitem[{Bennett and Wakefield(2001)}]{bennett_errors--variables_2001}
\bibinfo{author}{Bennett, J.}, \bibinfo{author}{Wakefield, J.}, \bibinfo{year}{2001}.
\newblock \bibinfo{title}{Errors-in-{Variables} in {Joint} {Population} {Pharmacokinetic}/{Pharmacodynamic} {Modeling}}.
\newblock \bibinfo{journal}{Biometrics} \bibinfo{volume}{57}, \bibinfo{pages}{803--812}.
\newblock \URLprefix \url{https://doi.org/10.1111/j.0006-341X.2001.00803.x}, \DOIprefix\doi{10.1111/j.0006-341X.2001.00803.x}. \bibinfo{note}{publisher: John Wiley \& Sons, Ltd}.
\bibitem[{Bradley et~al.(2015)Bradley, Holan and Wikle}]{Bradley2015multiscale}
\bibinfo{author}{Bradley, J.R.}, \bibinfo{author}{Holan, S.H.}, \bibinfo{author}{Wikle, C.K.}, \bibinfo{year}{2015}.
\newblock \bibinfo{title}{Multiscale spatial modeling of rates using the american community survey}.
\newblock \bibinfo{journal}{Journal of the Royal Statistical Society: Series A} \bibinfo{volume}{178}, \bibinfo{pages}{173--190}.
\bibitem[{{Center for Behavioral Health Statistics and Quality}(2021)}]{CBHSQ2021}
\bibinfo{author}{{Center for Behavioral Health Statistics and Quality}}, \bibinfo{year}{2021}.
\newblock \bibinfo{title}{2020 Methodological Summary and Definitions}.
\newblock \bibinfo{type}{Technical Report}. Substance Abuse and Mental Health Services Administration. \bibinfo{address}{Rockville, MD}.
\newblock \URLprefix \url{https://www.samhsa.gov/data/report/2020-methodological-summary-and-definitions}. \bibinfo{note}{accessed: April 20, 2025}.
\bibitem[{{Centers for Disease Control and Prevention (CDC)}(2023)}]{opioid_overdose}
\bibinfo{author}{{Centers for Disease Control and Prevention (CDC)}}, \bibinfo{year}{2023}.
\newblock \bibinfo{title}{Drug overdose deaths: Drug overdose deaths remained high in 2021}.
\newblock \bibinfo{howpublished}{\url{https://www.cdc.gov/drugoverdose/deaths/index.html}}.
\bibitem[{Dasgupta et~al.(2018)Dasgupta, Beletsky and Ciccarone}]{Dasgupta2018}
\bibinfo{author}{Dasgupta, N.}, \bibinfo{author}{Beletsky, L.}, \bibinfo{author}{Ciccarone, D.}, \bibinfo{year}{2018}.
\newblock \bibinfo{title}{Opioid epidemic: Addressing social determinants of health}.
\newblock \bibinfo{journal}{International Journal of Drug Policy} \bibinfo{volume}{59}, \bibinfo{pages}{1--2}.
\bibitem[{Gabry et~al.(2019)Gabry, Simpson, Vehtari, Betancourt and Gelman}]{gabry_visualization_2019}
\bibinfo{author}{Gabry, J.}, \bibinfo{author}{Simpson, D.}, \bibinfo{author}{Vehtari, A.}, \bibinfo{author}{Betancourt, M.}, \bibinfo{author}{Gelman, A.}, \bibinfo{year}{2019}.
\newblock \bibinfo{title}{Visualization in {Bayesian} workflow}.
\newblock \bibinfo{journal}{Journal of the Royal Statistical Society: Series A (Statistics in Society)} \bibinfo{volume}{182}, \bibinfo{pages}{389--402}.
\newblock \URLprefix \url{https://rss.onlinelibrary.wiley.com/doi/abs/10.1111/rssa.12378}, \DOIprefix\doi{10.1111/rssa.12378}. \bibinfo{note}{\_eprint: https://rss.onlinelibrary.wiley.com/doi/pdf/10.1111/rssa.12378}.
\bibitem[{Gao and Wakefield(2022)}]{gao2022spatial}
\bibinfo{author}{Gao, Y.}, \bibinfo{author}{Wakefield, J.}, \bibinfo{year}{2022}.
\newblock \bibinfo{title}{Spatial variance smoothing for small area estimation of demographic rates}.
\newblock \bibinfo{journal}{arXiv preprint arXiv:2209.02602} \URLprefix \url{https://arxiv.org/abs/2209.02602}.
\bibitem[{Ghosh and Rao(1994)}]{ghosh1994small}
\bibinfo{author}{Ghosh, M.}, \bibinfo{author}{Rao, J.N.K.}, \bibinfo{year}{1994}.
\newblock \bibinfo{title}{Small area estimation: An appraisal}.
\newblock \bibinfo{journal}{Statistical Science} \bibinfo{volume}{9}, \bibinfo{pages}{55--76}.
\bibitem[{Griffin(2020)}]{griffin_spatial_2020}
\bibinfo{author}{Griffin, S.}, \bibinfo{year}{2020}.
\newblock \bibinfo{title}{Spatial downscaling disease risk using random forests machine learning}.
\newblock \bibinfo{type}{Technical Report}. Engineer Research and Development Center (U.S.).
\newblock \URLprefix \url{https://hdl.handle.net/11681/35618}, \DOIprefix\doi{10.21079/11681/35618}.
\bibitem[{Hastie and Tibshirani(2009)}]{hastie_model_2009}
\bibinfo{author}{Hastie, T.}, \bibinfo{author}{Tibshirani, R.}, \bibinfo{year}{2009}.
\newblock \bibinfo{title}{Model {Robustness} and {Validation} in {Predictive} {Modeling}}.
\newblock \bibinfo{journal}{Statistical Science} \bibinfo{volume}{24}, \bibinfo{pages}{570--586}.
\newblock \DOIprefix\doi{10.1214/09-STS297}.
\bibitem[{{Hepler SA} et~al.(2023){Hepler SA}, {Kline DM}, {Bonny A}, {McKnight E} and {Waller LA}}]{hepler_integrated_2023}
\bibinfo{author}{{Hepler SA}}, \bibinfo{author}{{Kline DM}}, \bibinfo{author}{{Bonny A}}, \bibinfo{author}{{McKnight E}}, \bibinfo{author}{{Waller LA}}, \bibinfo{year}{2023}.
\newblock \bibinfo{title}{An integrated abundance model for estimating county-level prevalence of opioid misuse in {Ohio}}.
\newblock \bibinfo{journal}{Journal of the Royal Statistical Society Series A: Statistics in Society} \bibinfo{volume}{186}, \bibinfo{pages}{43--60}.
\newblock \bibinfo{note}{Publisher: Oxford University Press US}.
\bibitem[{Hogg et~al.(2023)Hogg, Quick and Wakefield}]{hogg2023two}
\bibinfo{author}{Hogg, R.}, \bibinfo{author}{Quick, H.}, \bibinfo{author}{Wakefield, J.}, \bibinfo{year}{2023}.
\newblock \bibinfo{title}{A two-stage bayesian small area estimation method for proportions with sparse data}.
\newblock \bibinfo{journal}{arXiv preprint arXiv:2306.11302} \URLprefix \url{https://arxiv.org/abs/2306.11302}.
\bibitem[{Jacob et~al.(2017)Jacob, Murray, Holmes and Robert}]{jacob_better_2017}
\bibinfo{author}{Jacob, P.E.}, \bibinfo{author}{Murray, L.M.}, \bibinfo{author}{Holmes, C.C.}, \bibinfo{author}{Robert, C.P.}, \bibinfo{year}{2017}.
\newblock \bibinfo{title}{Better together? {Statistical} learning in models made of modules}.
\newblock \bibinfo{journal}{arXiv preprint arXiv:1708.08719} .
\bibitem[{Kline et~al.(2021)Kline, Ji and Hepler}]{kline2021multivariate}
\bibinfo{author}{Kline, D.}, \bibinfo{author}{Ji, Y.}, \bibinfo{author}{Hepler, S.}, \bibinfo{year}{2021}.
\newblock \bibinfo{title}{A multivariate spatio-temporal model of the opioid epidemic in ohio: a factor model approach}.
\newblock \bibinfo{journal}{Health Services and Outcomes Research Methodology} \bibinfo{volume}{21}, \bibinfo{pages}{42--53}.
\bibitem[{{Kline D} and {Hepler SA}(2021)}]{kline_estimating_2021}
\bibinfo{author}{{Kline D}}, \bibinfo{author}{{Hepler SA}}, \bibinfo{year}{2021}.
\newblock \bibinfo{title}{Estimating the burden of the opioid epidemic for adults and adolescents in {Ohio} counties}.
\newblock \bibinfo{journal}{Biometrics} \bibinfo{volume}{77}, \bibinfo{pages}{765--775}.
\newblock \bibinfo{note}{Publisher: Wiley Online Library}.
\bibitem[{{Kline D} et~al.(2023){Kline D}, {Waller LA}, {McKnight E}, {Bonny A}, {Miller WC} and {Hepler SA}}]{kline_dynamic_2023}
\bibinfo{author}{{Kline D}}, \bibinfo{author}{{Waller LA}}, \bibinfo{author}{{McKnight E}}, \bibinfo{author}{{Bonny A}}, \bibinfo{author}{{Miller WC}}, \bibinfo{author}{{Hepler SA}}, \bibinfo{year}{2023}.
\newblock \bibinfo{title}{A {Dynamic} {Spatial} {Factor} {Model} to {Describe} the {Opioid} {Syndemic} in {Ohio}}.
\newblock \bibinfo{journal}{Epidemiology} \bibinfo{volume}{34}, \bibinfo{pages}{487--494}.
\newblock \bibinfo{note}{Publisher: Wolters Kluwer}.
\bibitem[{Konstantinoudis et~al.(2021)Konstantinoudis, Padellini, Bennett, Davies, Ezzati and Blangiardo}]{konstantinoudis_long-term_2021}
\bibinfo{author}{Konstantinoudis, G.}, \bibinfo{author}{Padellini, T.}, \bibinfo{author}{Bennett, J.}, \bibinfo{author}{Davies, B.}, \bibinfo{author}{Ezzati, M.}, \bibinfo{author}{Blangiardo, M.}, \bibinfo{year}{2021}.
\newblock \bibinfo{title}{Long-term exposure to air-pollution and {COVID}-19 mortality in {England}: {A} hierarchical spatial analysis}.
\newblock \bibinfo{journal}{Environment International} \bibinfo{volume}{146}, \bibinfo{pages}{106316}.
\newblock \URLprefix \url{https://linkinghub.elsevier.com/retrieve/pii/S0160412020322716}, \DOIprefix\doi{10.1016/j.envint.2020.106316}.
\bibitem[{Linard and Tatem(2012)}]{linard_large-scale_2012}
\bibinfo{author}{Linard, C.}, \bibinfo{author}{Tatem, A.J.}, \bibinfo{year}{2012}.
\newblock \bibinfo{title}{Large-scale spatial population databases in infectious disease research}.
\newblock \bibinfo{journal}{International Journal of Health Geographics} \bibinfo{volume}{11}, \bibinfo{pages}{7}.
\newblock \URLprefix \url{http://ij-healthgeographics.biomedcentral.com/articles/10.1186/1476-072X-11-7}, \DOIprefix\doi{10.1186/1476-072X-11-7}.
\bibitem[{Mercer et~al.(2015)Mercer, Wakefield, Chen, Lumley et~al.}]{mercer2015small}
\bibinfo{author}{Mercer, L.D.}, \bibinfo{author}{Wakefield, J.}, \bibinfo{author}{Chen, D.}, \bibinfo{author}{Lumley, T.}, et~al., \bibinfo{year}{2015}.
\newblock \bibinfo{title}{Small area estimation of child mortality in the absence of vital registration}.
\newblock \bibinfo{journal}{Annals of Applied Statistics} \bibinfo{volume}{9}, \bibinfo{pages}{1889--1905}.
\newblock \DOIprefix\doi{10.1214/15-AOAS872}.
\bibitem[{Monnat(2018)}]{Monnat2018}
\bibinfo{author}{Monnat, S.M.}, \bibinfo{year}{2018}.
\newblock \bibinfo{title}{Factors associated with county-level differences in u.s. drug-related mortality rates}.
\newblock \bibinfo{journal}{American Journal of Preventive Medicine} \bibinfo{volume}{54}, \bibinfo{pages}{611--619}.
\bibitem[{Nandi et~al.(2020)Nandi, Lucas, Arambepola, Gething and Weiss}]{nandi_disaggregation_2020}
\bibinfo{author}{Nandi, A.K.}, \bibinfo{author}{Lucas, T.C.D.}, \bibinfo{author}{Arambepola, R.}, \bibinfo{author}{Gething, P.}, \bibinfo{author}{Weiss, D.J.}, \bibinfo{year}{2020}.
\newblock \bibinfo{title}{disaggregation: {An} {R} {Package} for {Bayesian} {Spatial} {Disaggregation} {Modelling}}.
\newblock \URLprefix \url{http://arxiv.org/abs/2001.04847}, \DOIprefix\doi{10.48550/arXiv.2001.04847}. \bibinfo{note}{arXiv:2001.04847 [stat]}.
\bibitem[{{National Institute on Drug Abuse}(2023a)}]{NIDA_OUD_2023}
\bibinfo{author}{{National Institute on Drug Abuse}}, \bibinfo{year}{2023}a.
\newblock \bibinfo{title}{Only 1 in 5 u.s. adults with opioid use disorder received medications to treat it in 2021}.
\newblock \URLprefix \url{https://nida.nih.gov/news-events/news-releases/2023/08/only-1-in-5-us-adults-with-opioid-use-disorder-received-medications-to-treat-it-in-2021}. \bibinfo{note}{accessed: 2025-01-21}.
\bibitem[{{National Institute on Drug Abuse}(2023b)}]{NIDA_OUD_2023A}
\bibinfo{author}{{National Institute on Drug Abuse}}, \bibinfo{year}{2023}b.
\newblock \bibinfo{title}{Opioid use disorder: What is oud?}
\newblock \URLprefix \url{https://nida.nih.gov/research-topics/opioids/opioid-use-disorder}.
\bibitem[{{Perry de Valpine} et~al.(){Perry de Valpine}, {Daniel Turek}, {Christopher Paciorek}, {Cliff Anderson-Bergman}, {Duncan Temple Lang} and {Ras Bodik}}]{perry_de_valpine_programming_nodate}
\bibinfo{author}{{Perry de Valpine}}, \bibinfo{author}{{Daniel Turek}}, \bibinfo{author}{{Christopher Paciorek}}, \bibinfo{author}{{Cliff Anderson-Bergman}}, \bibinfo{author}{{Duncan Temple Lang}}, \bibinfo{author}{{Ras Bodik}}, .
\newblock \bibinfo{title}{Programming with models: {Writing} statistical algorithms for general model structures with {NIMBLE}}.
\newblock \DOIprefix\doi{10.1080/10618600.2016.1172487}.
\bibitem[{Peterson et~al.(2024)Peterson, Guranich, Cresswell and Alkema}]{peterson2024bayesian}
\bibinfo{author}{Peterson, E.N.}, \bibinfo{author}{Guranich, G.}, \bibinfo{author}{Cresswell, J.A.}, \bibinfo{author}{Alkema, L.}, \bibinfo{year}{2024}.
\newblock \bibinfo{title}{A bayesian approach to estimate maternal mortality globally using national civil registration vital statistics data accounting for reporting errors}.
\newblock \bibinfo{journal}{Statistics and Public Policy} \bibinfo{volume}{11}, \bibinfo{pages}{2286313}.
\bibitem[{Peterson et~al.(2023)Peterson, Nethery, Chen, Tabb, Coull, Piel and Waller}]{peterson2023bayesian}
\bibinfo{author}{Peterson, E.N.}, \bibinfo{author}{Nethery, R.C.}, \bibinfo{author}{Chen, J.T.}, \bibinfo{author}{Tabb, L.P.}, \bibinfo{author}{Coull, B.A.}, \bibinfo{author}{Piel, F.B.}, \bibinfo{author}{Waller, L.A.}, \bibinfo{year}{2023}.
\newblock \bibinfo{title}{A bayesian spatial berkson error approach to estimate small area opioid mortality rates accounting for population-at-risk uncertainty}.
\newblock \bibinfo{journal}{arXiv preprint arXiv:2312.13331} .
\bibitem[{Plummer(2015)}]{plummer_cuts_2015}
\bibinfo{author}{Plummer, M.}, \bibinfo{year}{2015}.
\newblock \bibinfo{title}{Cuts in {Bayesian} graphical models}.
\newblock \bibinfo{journal}{Statistics and Computing} \bibinfo{volume}{25}, \bibinfo{pages}{37--43}.
\newblock \URLprefix \url{https://doi.org/10.1007/s11222-014-9503-z}, \DOIprefix\doi{10.1007/s11222-014-9503-z}.
\bibitem[{Plummer(2017)}]{plummer_jags_2017}
\bibinfo{author}{Plummer, M.}, \bibinfo{year}{2017}.
\newblock \bibinfo{title}{{JAGS}: {A} program for analysis of {Bayesian} graphical models using {Gibbs} sampling}.
\bibitem[{{Population Estimation Program, U.S. Census Bureau}(2019)}]{pep}
\bibinfo{author}{{Population Estimation Program, U.S. Census Bureau}}, \bibinfo{year}{2019}.
\newblock \bibinfo{title}{Methodology for the united states population estimates: Vintage 2019}.
\newblock \bibinfo{howpublished}{\url{https://www2.census.gov/programs-surveys/popest/technical-documentation/methodology/2010-2019/natstcopr-methv2.pdf}}.
\newblock \URLprefix \url{https://www2.census.gov/programs-surveys/popest/technical-documentation/methodology/2010-2019/natstcopr-methv2.pdf}. \bibinfo{note}{accessed on 07-05-2021}.
\bibitem[{Python et~al.(2022)Python, Bender, Blangiardo, Illian, Lin, Liu, Lucas, Tan, Wen, Svanidze and Yin}]{python_downscaling_2022}
\bibinfo{author}{Python, A.}, \bibinfo{author}{Bender, A.}, \bibinfo{author}{Blangiardo, M.}, \bibinfo{author}{Illian, J.B.}, \bibinfo{author}{Lin, Y.}, \bibinfo{author}{Liu, B.}, \bibinfo{author}{Lucas, T.C.}, \bibinfo{author}{Tan, S.}, \bibinfo{author}{Wen, Y.}, \bibinfo{author}{Svanidze, D.}, \bibinfo{author}{Yin, J.}, \bibinfo{year}{2022}.
\newblock \bibinfo{title}{A {Downscaling} {Approach} to {Compare} {COVID}-19 {Count} {Data} from {Databases} {Aggregated} at {Different} {Spatial} {Scales}}.
\newblock \bibinfo{journal}{Journal of the Royal Statistical Society Series A: Statistics in Society} \bibinfo{volume}{185}, \bibinfo{pages}{202--218}.
\newblock \URLprefix \url{https://academic.oup.com/jrsssa/article/185/1/202/7068448}, \DOIprefix\doi{10.1111/rssa.12738}.
\bibitem[{Rao and Molina(2015)}]{rao2015small}
\bibinfo{author}{Rao, J.N.K.}, \bibinfo{author}{Molina, I.}, \bibinfo{year}{2015}.
\newblock \bibinfo{title}{Small Area Estimation}.
\newblock \bibinfo{publisher}{Wiley}, \bibinfo{address}{Hoboken, NJ}.
\bibitem[{Rosenblum and Unick(2022)}]{Rosenblum2022}
\bibinfo{author}{Rosenblum, D.}, \bibinfo{author}{Unick, G.J.}, \bibinfo{year}{2022}.
\newblock \bibinfo{title}{The role of health systems in addressing the opioid overdose crisis}.
\newblock \bibinfo{journal}{Annual Review of Public Health} \bibinfo{volume}{43}, \bibinfo{pages}{311--329}.
\bibitem[{Rue et~al.(2009)Rue, Martino and Chopin}]{rue2009inla}
\bibinfo{author}{Rue, H.}, \bibinfo{author}{Martino, S.}, \bibinfo{author}{Chopin, N.}, \bibinfo{year}{2009}.
\newblock \bibinfo{title}{Approximate bayesian inference for latent gaussian models by using integrated nested laplace approximations}.
\newblock \bibinfo{journal}{Journal of the Royal Statistical Society: Series B (Statistical Methodology)} \bibinfo{volume}{71}, \bibinfo{pages}{319--392}.
\newblock \DOIprefix\doi{10.1111/j.1467-9868.2008.00700.x}.
\bibitem[{Rue et~al.(2024)}]{r-inla}
\bibinfo{author}{Rue, H.}, et~al., \bibinfo{year}{2024}.
\newblock \bibinfo{title}{R-inla project}.
\newblock \bibinfo{note}{\url{https://www.r-inla.org}}.
\bibitem[{{Rue, H.} and {Held, L.}(2005)}]{havard_rue_gaussian_2005}
\bibinfo{author}{{Rue, H.}}, \bibinfo{author}{{Held, L.}}, \bibinfo{year}{2005}.
\newblock \bibinfo{title}{Gaussian {Markov} {Random} {Fields}}.
\newblock \bibinfo{publisher}{Chapman \& Hall/CRC}.
\bibitem[{{Substance Abuse and Mental Health Services Administration}(2023)}]{NSDUH_OUD_2023}
\bibinfo{author}{{Substance Abuse and Mental Health Services Administration}}, \bibinfo{year}{2023}.
\newblock \bibinfo{title}{2021 national survey on drug use and health}.
\newblock \URLprefix \url{https://www.samhsa.gov/data/report/2021-nsduh-detailed-tables}.
\bibitem[{Ugarte et~al.(2020)Ugarte, Goicoa and Militino}]{Ugarte2020benchmarking}
\bibinfo{author}{Ugarte, M.D.}, \bibinfo{author}{Goicoa, T.}, \bibinfo{author}{Militino, A.F.}, \bibinfo{year}{2020}.
\newblock \bibinfo{title}{Bayesian benchmarking techniques in small area estimation}.
\newblock \bibinfo{journal}{Statistical Science} \bibinfo{volume}{35}, \bibinfo{pages}{261--284}.
\bibitem[{{United States Department of Health and Human Services}(2016)}]{NSDUH2014}
\bibinfo{author}{{United States Department of Health and Human Services}, {Substance Abuse and Mental Health Services Administration}, C.}, \bibinfo{year}{2016}.
\newblock \bibinfo{title}{{National Survey on Drug Use and Health, 2014}}.
\newblock \URLprefix \url{https://doi.org/10.3886/ICPSR36361.v1}, \DOIprefix\doi{10.3886/ICPSR36361.v1}. \bibinfo{note}{public Use File}.
\bibitem[{{United States Department of Health and Human Services}(2023)}]{NSDUH2022}
\bibinfo{author}{{United States Department of Health and Human Services}, {Substance Abuse and Mental Health Services Administration}, C.}, \bibinfo{year}{2023}.
\newblock \bibinfo{title}{{National Survey on Drug Use and Health, 2022}}.
\newblock \URLprefix \url{https://www.samhsa.gov/data/system/files/media-puf-file/NSDUH-2022-DS0001-info-codebook.pdf}. \bibinfo{note}{public Use File}.
\bibitem[{{U.S. Census Bureau}(2025)}]{USCensusPEP}
\bibinfo{author}{{U.S. Census Bureau}}, \bibinfo{year}{2025}.
\newblock \bibinfo{title}{{Population and Housing Unit Estimates}}.
\newblock \URLprefix \url{https://www.census.gov/popest}. \bibinfo{note}{population Estimates Program (PEP), Vintage 2024}.
\bibitem[{{U.S. Department of Health and Human Services}(2020a)}]{HHS_Opioid_2020}
\bibinfo{author}{{U.S. Department of Health and Human Services}}, \bibinfo{year}{2020}a.
\newblock \bibinfo{title}{Advancing the collection and use of opioid-related data: Progress and opportunities}.
\newblock \URLprefix \url{https://aspe.hhs.gov/sites/default/files/private/pdf/259016/NORC-2020-ASPE-Opi-Vignette.pdf}. \bibinfo{note}{accessed: 2025-01-21}.
\bibitem[{{U.S. Department of Health and Human Services}(2020b)}]{HHS_Opioid_2020A}
\bibinfo{author}{{U.S. Department of Health and Human Services}}, \bibinfo{year}{2020}b.
\newblock \bibinfo{title}{Hhs strategy to combat the opioid crisis}.
\newblock \URLprefix \url{https://www.hhs.gov/opioids/about-the-epidemic/index.html}.
\bibitem[{{Vehtari, A.} et~al.(2021){Vehtari, A.}, {Gelman, A.}, {Simpson, D.}, {Carpenter, B.} and {Bürkner, P.C.}}]{aki_vehtari_rank-normalization_2021}
\bibinfo{author}{{Vehtari, A.}}, \bibinfo{author}{{Gelman, A.}}, \bibinfo{author}{{Simpson, D.}}, \bibinfo{author}{{Carpenter, B.}}, \bibinfo{author}{{Bürkner, P.C.}}, \bibinfo{year}{2021}.
\newblock \bibinfo{title}{Rank-{Normalization}, {Folding}, and {Localization}: {An} {Improved} \${\textbackslash}widehat\{{R}\}\$ for {Assessing} {Convergence} of {MCMC} (with {Discussion})}.
\newblock \bibinfo{journal}{Bayesian Analysis} \bibinfo{volume}{16}, \bibinfo{pages}{667--718}.
\newblock \URLprefix \url{https://doi.org/10.1214/20-BA1221}, \DOIprefix\doi{10.1214/20-BA1221}.
\bibitem[{{Wakefield, J.}(2007)}]{jon_wakefield_disease_2007}
\bibinfo{author}{{Wakefield, J.}}, \bibinfo{year}{2007}.
\newblock \bibinfo{title}{Disease mapping and spatial regression with count data}.
\newblock \bibinfo{journal}{Biostatistics} \bibinfo{volume}{8}, \bibinfo{pages}{158--183}.
\newblock \DOIprefix\doi{10.1093/biostatistics/kx1008}.
\bibitem[{{Walker, K.}(2020)}]{walker_k_tidycensus_2020}
\bibinfo{author}{{Walker, K.}}, \bibinfo{year}{2020}.
\newblock \bibinfo{title}{tidycensus: {Load} {US} {Census} {Boundary} and {Attribute} {Data} as `tidyverse'. {R} package version 0.9.9.2}.
\newblock \URLprefix \url{https://walker-data.com/tidycensus/articles/basic-usage.html}.
\bibitem[{{Waller, L.A.} and {Gotway, C.A.}(2004)}]{waller_gotway}
\bibinfo{author}{{Waller, L.A.}}, \bibinfo{author}{{Gotway, C.A.}}, \bibinfo{year}{2004}.
\newblock \bibinfo{title}{Applied Spatial Statistics for Public Health Data}.
\newblock \bibinfo{publisher}{John Wiley \& Sons, Inc.}
\newblock \DOIprefix\doi{10.1002/0471662682}.
\bibitem[{Wulandari et~al.(2023)Wulandari, Best and Richardson}]{wulandari2023overdispersion}
\bibinfo{author}{Wulandari, E.}, \bibinfo{author}{Best, N.}, \bibinfo{author}{Richardson, S.}, \bibinfo{year}{2023}.
\newblock \bibinfo{title}{Addressing overdispersion in small-area count data: a bayesian perspective}.
\newblock \bibinfo{journal}{Biostatistics} \bibinfo{volume}{24}, \bibinfo{pages}{383--401}.
\bibitem[{You and Rao(2020)}]{census2020}
\bibinfo{author}{You, H.}, \bibinfo{author}{Rao, J.N.K.}, \bibinfo{year}{2020}.
\newblock \bibinfo{title}{Bayesian Hierarchical Spatial Models for Small Area Estimation}.
\newblock \bibinfo{type}{Technical Report}. U.S. Census Bureau.
\newblock \URLprefix \url{https://www.census.gov/library/working-papers/2020/adrm/RRS2020-07.html}.

\end{thebibliography}
\bibliographystyle{elsarticle-harv}

\clearpage

\appendix
\section{Temporal Covariate Trends at the State and County Level}\label{sec:covs}
\begin{figure}[ht]
  \centering
  \includegraphics[width=0.7\textwidth]{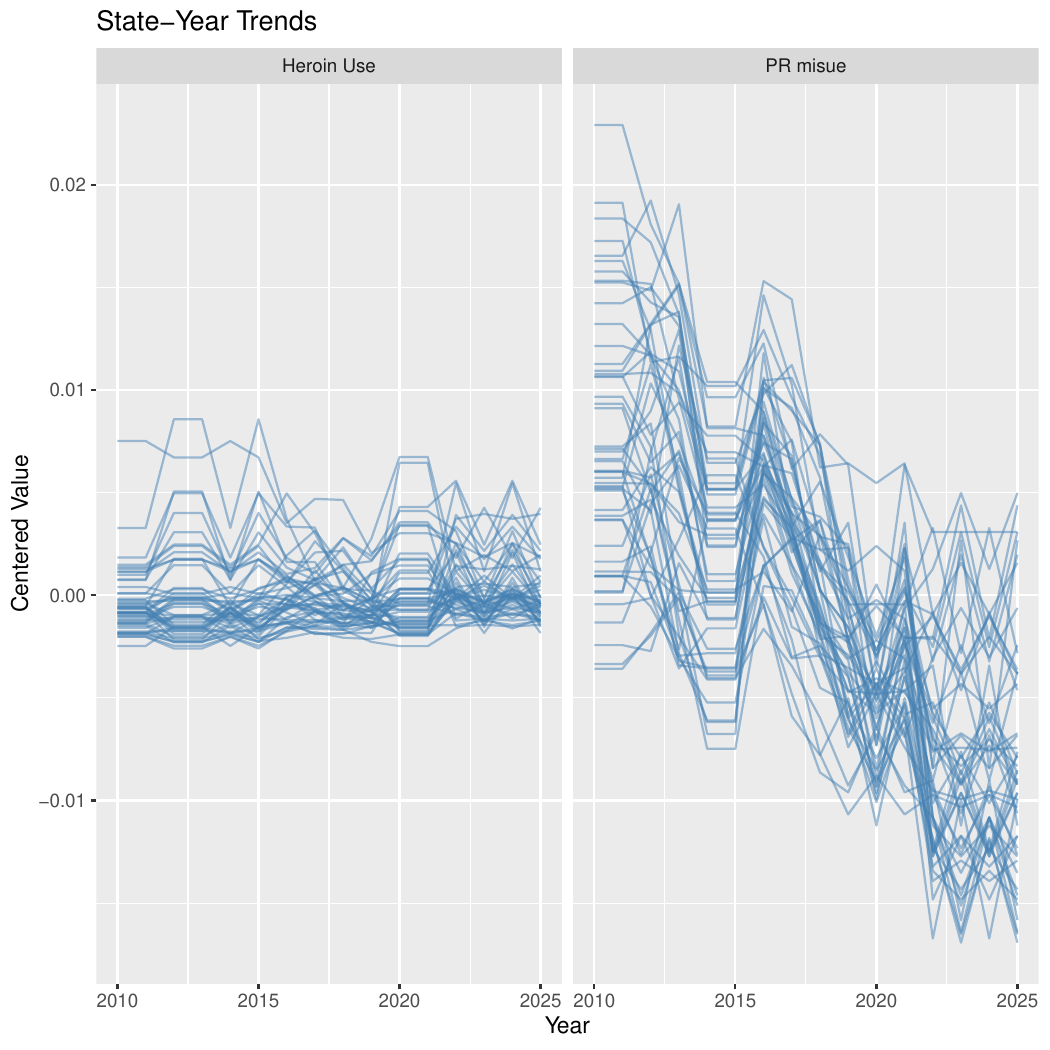}
  \caption{State-level covariate trends for heroin use (left) and PR misuse (right). Lines denote different states. }
  \label{fig:state_covs}
\end{figure}

Figure~\ref{fig:state_covs} displays state-level trends in heroin use and prescription pain reliever (PR) misuse, with each line representing a different state. These covariates are derived from national survey data and reflect variation in substance use patterns across states from 2010 to 2025. Values for 2024-2025 were interpolated using linear interpolation methods. 

The following pages present county-level covariate trends, disaggregated by state and grouped by variable. Each line represents a county, and each panel corresponds to a specific covariate (e.g., opioid-related mortality, rurality, poverty rate, disability rate, and cumulative opioid prescribing). Covariates have been centered on the national mean to facilitate interpretability. States are split into manageable groups across pages to improve readability.

\includepdf[pages=-, scale=0.7]{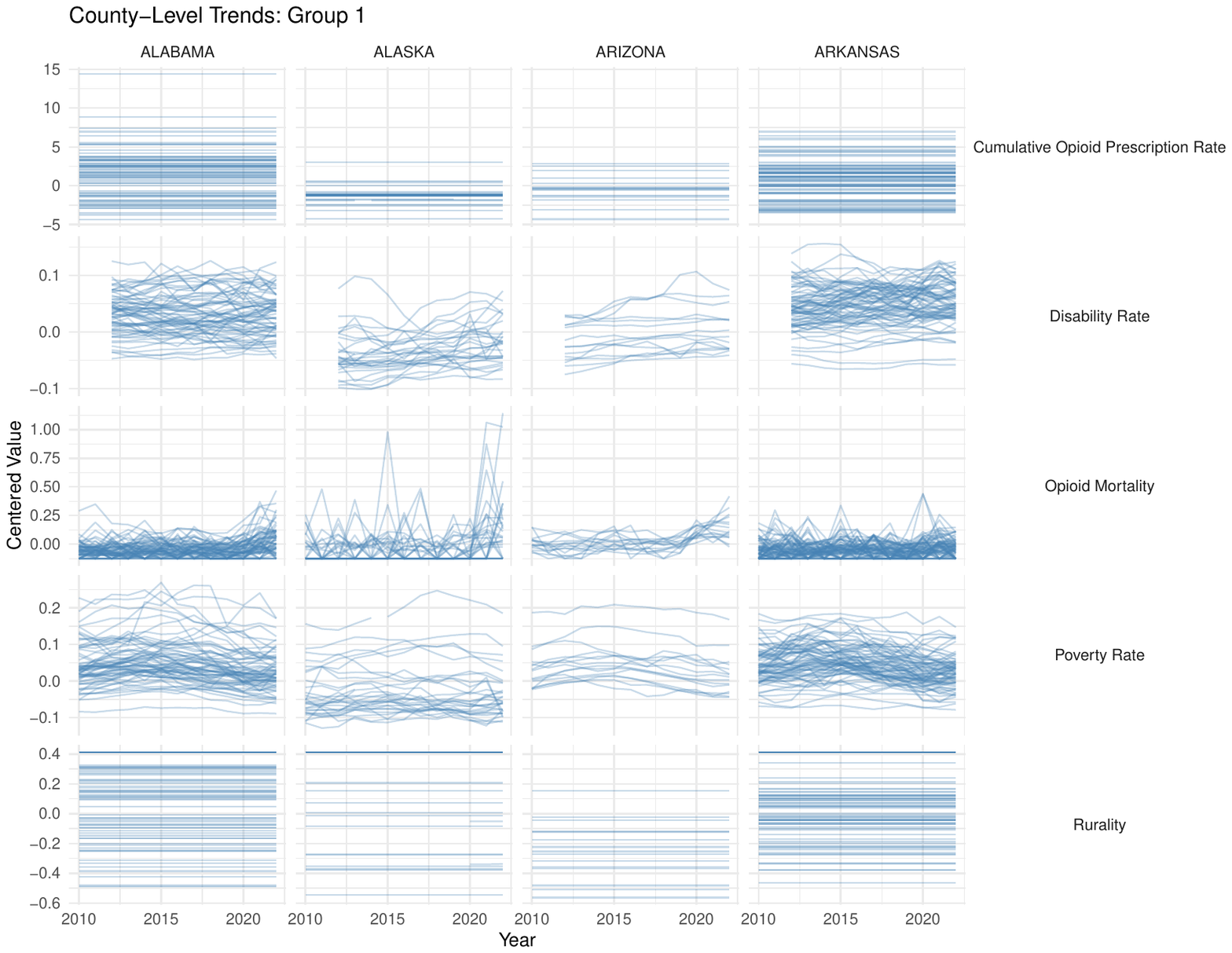}

\section{Notation Summary}\label{sec:A}
\begin{table}[ht]
\centering
\caption{Summary of model notation for Stage I (state-level) and Stage II (county-level).}
\renewcommand{\arraystretch}{1.3}
\begin{tabular}{|l|p{12cm}|}
\hline
\multicolumn{2}{|c|}{\textbf{Stage I: State-Level Model}} \\
\hline
$y_{s,t}$ & Observed OUD case count in state $s$ and year $t$ \\
$n_{s,t}$ & Population at risk in state $s$ and year $t$ \\
$\pi_{s,t}$ & True state-level OUD incidence risk \\
$\tilde{\pi}_{s,t}$ & Posterior estimate of $\pi_{s,t}$ from Stage I \\
$\lambda_{s,t}$ & Latent log-risk for state $s$, year $t$: $\lambda_{s,t} = \log(\pi_{s,t})$ \\
$\tilde{y}_{s,t}$ & Posterior estimate of state-level count: $\tilde{y}_{s,t} = \tilde{\pi}_{s,t} \cdot n_{s,t}$ \\
$\alpha$ & Global intercept in the state-level model \\
$\beta_j$ & Regression coefficient for covariate $x_{j,s,t}$ \\
$x_{j,s,t}$ & Covariate $j$ for state $s$ and year $t$ \\
$\omega_s$ & Random intercept for state $s$ \\
$\tilde{\omega}_s$ & Posterior estimate of $\omega_s$ passed to Stage II \\
$\phi_{s,t}$ & Temporally structured random effect (first-order random walk) \\
$\Delta_{s,t}$ & Adjustment factor for case definition change post-2020 \\
$r_{s,t}$ & Lower bound on $\Delta_{s,t}$, derived from 2016-to-current ratios \\
\hline
\multicolumn{2}{|c|}{\textbf{Stage II: County-Level Model}} \\
\hline
$\mu_{c,t}$ & Latent OUD case count in county $c$ and year $t$ \\
$n_{c,t}$ & Population at risk in county $c$ and year $t$ \\
$\pi_{c,t}$ & True county-level OUD incidence risk \\
$\tilde{\pi}_{c,t}$ & Posterior estimate of county-level risk from Stage II \\
$\mu_{c,t}$ & Expected count in county $c$, year $t$: $\mu_{c,t} = \rho_{c,t} \cdot \tilde{y}_{s,t}$ \\
$\rho_{c,t}$ & Softmax-normalized probability of a case in county $c$ within state $s$ \\
$\eta_{c,t}$ & Linear predictor: $\eta_{c,t} = \alpha + \sum_j \beta_j x_{j,c,t} + \phi_{c,t} + \tilde{\omega}_s$ \\
$x_{j,c,t}$ & Covariate $j$ for county $c$ and year $t$ \\
$\beta_j$ & Regression coefficient for county-level covariate $x_{j,c,t}$ \\
$f_{\text{county}}(c)$ & Spatial random effect modeled using BYM or ICAR structure \\
$\delta_{c,t}$ & Temporally structured deviation for county $c$ (RW1 process) \\
$\phi_{c,t}$ & Combined spatio-temporal random effect: $\phi_{c,t} = f_{\text{county}}(c) + \delta_{c,t}$ \\
$\alpha$ & Global intercept for the county-level model \\
\hline
\end{tabular}
\label{tab:notation_summary}
\end{table}

\section{State Model: Posterior Inference and Full Conditional Distributions}\label{sec:posts}

Let $\boldsymbol{y} = \{y_i\}_{i=1}^N$ denote the vector of observed OUD case counts across all state-year combinations, and let $\boldsymbol{n} = \{n_i\}_{i=1}^N$ denote the corresponding population denominators. The full set of unknown parameters is given by:

\[
\boldsymbol{\Theta} = \left\{\alpha, \beta_1, \beta_2, \omega_s, \phi_{s,t}, \pi_{s,t}, \lambda_{s,t}, \Delta_{s,t} \right\}.
\]

The joint posterior distribution is proportional to the product of the likelihood and the prior distributions:

\begin{align}
p(\boldsymbol{\Theta} \mid \boldsymbol{y}, \boldsymbol{n}) &\propto \underbrace{\prod_{i=1}^{N} \text{Poisson}\left(y_i \,\middle|\, \frac{n_i \cdot \pi_{s[i],t[i]}}{\Delta_{s[i],t[i]}} \right)}_{\text{Likelihood}} \times
\underbrace{\prod_{s,t} p(\pi_{s,t} \mid \lambda_{s,t})}_{\text{Latent process}} \times
\underbrace{p(\lambda_{s,t} \mid \alpha, \boldsymbol{\beta}, \omega_s, \phi_{s,t})}_{\text{Linear predictor}} \nonumber \\
&\quad \times \underbrace{\prod_s p(\omega_s)}_{\text{Random intercept}} \times 
\underbrace{\prod_{s,t} p(\phi_{s,t} \mid \phi_{s,t\pm1}, \tau_\phi)}_{\text{Random walk}} \times 
\underbrace{\prod_{s,t>t_0} p(\Delta_{s,t} \mid r_{s,t})}_{\text{Adjustment factors}} \times 
p(\alpha) \prod_j p(\beta_j) \times p(\tau_\phi).
\end{align}

\paragraph{Full Conditional for \texorpdfstring{$\pi_{s,t}$}{pi\_st}.}

The latent risk $\pi_{s,t}$ is modeled on the log scale as:
\[
\log(\pi_{s,t}) = \lambda_{s,t}, \quad \text{so that } \pi_{s,t} = \exp(\lambda_{s,t}).
\]

Thus, the posterior distribution of $\pi_{s,t}$ is obtained via transformation of the posterior samples of $\lambda_{s,t}$:

\[
p(\pi_{s,t} \mid \text{data}) = \int p(\pi_{s,t} \mid \lambda_{s,t}) \cdot p(\lambda_{s,t} \mid \text{data}) \, d\lambda_{s,t},
\]

where:
\[
\pi_{s,t} = \exp(\lambda_{s,t}) \quad \text{and} \quad p(\lambda_{s,t} \mid \text{data}) \text{ is sampled via MCMC}.
\]

\paragraph{Full Conditional for \texorpdfstring{$\lambda_{s,t}$}{lambda\_st}.}

The full conditional distribution for $\lambda_{s,t}$ is given by:

\[
p(\lambda_{s,t} \mid y_i, \boldsymbol{\Theta}_{- \lambda_{s,t}}) \propto \prod_{i \in \mathcal{I}_{s,t}} \text{Poisson}\left(y_i \,\middle|\, \frac{n_i \cdot e^{\lambda_{s,t}}}{\Delta_{s,t}} \right) \cdot \mathcal{N}(\lambda_{s,t} \mid \alpha + \beta_1 x_{1,s,t} + \beta_2 x_{2,s,t} + \omega_s + \phi_{s,t}, \sigma^2_{\lambda}),
\]

where $\mathcal{I}_{s,t}$ indexes all observations from state $s$ and year $t$, and $\sigma^2_\lambda$ is implicitly defined by the Gaussian priors.

\paragraph{Full Conditional for \texorpdfstring{$\Delta_{s,t}$}{Delta\_st}.}

For $t > t_0$, the adjustment factor $\Delta_{s,t}$ has a truncated uniform prior:

\[
\Delta_{s,t} \sim \text{Uniform}(r_{s,t}, 1),
\]

and the corresponding full conditional is:

\[
p(\Delta_{s,t} \mid y_i, \pi_{s,t}, n_i) \propto \prod_{i \in \mathcal{I}_{s,t}} \text{Poisson}\left(y_i \,\middle|\, \frac{n_i \cdot \pi_{s,t}}{\Delta_{s,t}} \right) \cdot \mathbb{I}[r_{s,t} \leq \Delta_{s,t} \leq 1].
\]

\paragraph{Remaining Full Conditionals.}

The full conditionals for $\alpha$, $\beta_j$, $\omega_s$, and $\phi_{s,t}$ follow standard forms under the Gaussian prior assumptions. For example:

\[
p(\omega_s \mid \cdot) \propto \prod_{t} \mathcal{N}(\lambda_{s,t} \mid \cdots + \omega_s + \cdots, \sigma^2) \cdot \mathcal{N}(\omega_s \mid 0, \tau_\omega^{-1}),
\]

\[
p(\phi_{s,t} \mid \phi_{s,t-1}, \phi_{s,t+1}, \tau_\phi) \propto \mathcal{N}(\lambda_{s,t} \mid \cdots + \phi_{s,t} + \cdots, \sigma^2) \cdot \mathcal{N}(\phi_{s,t} \mid \phi_{s,t-1}, \tau_\phi^{-1}) \cdot \mathcal{N}(\phi_{s,t+1} \mid \phi_{s,t}, \tau_\phi^{-1}).
\]

Posterior summaries for parameters of interest, such as $\pi_{s,t}$, are obtained by transforming posterior draws of $\lambda_{s,t}$ and reporting posterior medians and 95\% credible intervals.

\section{State Trends in OUD}
\begin{figure}[H]
  \centering
  \includegraphics[width=\textwidth]{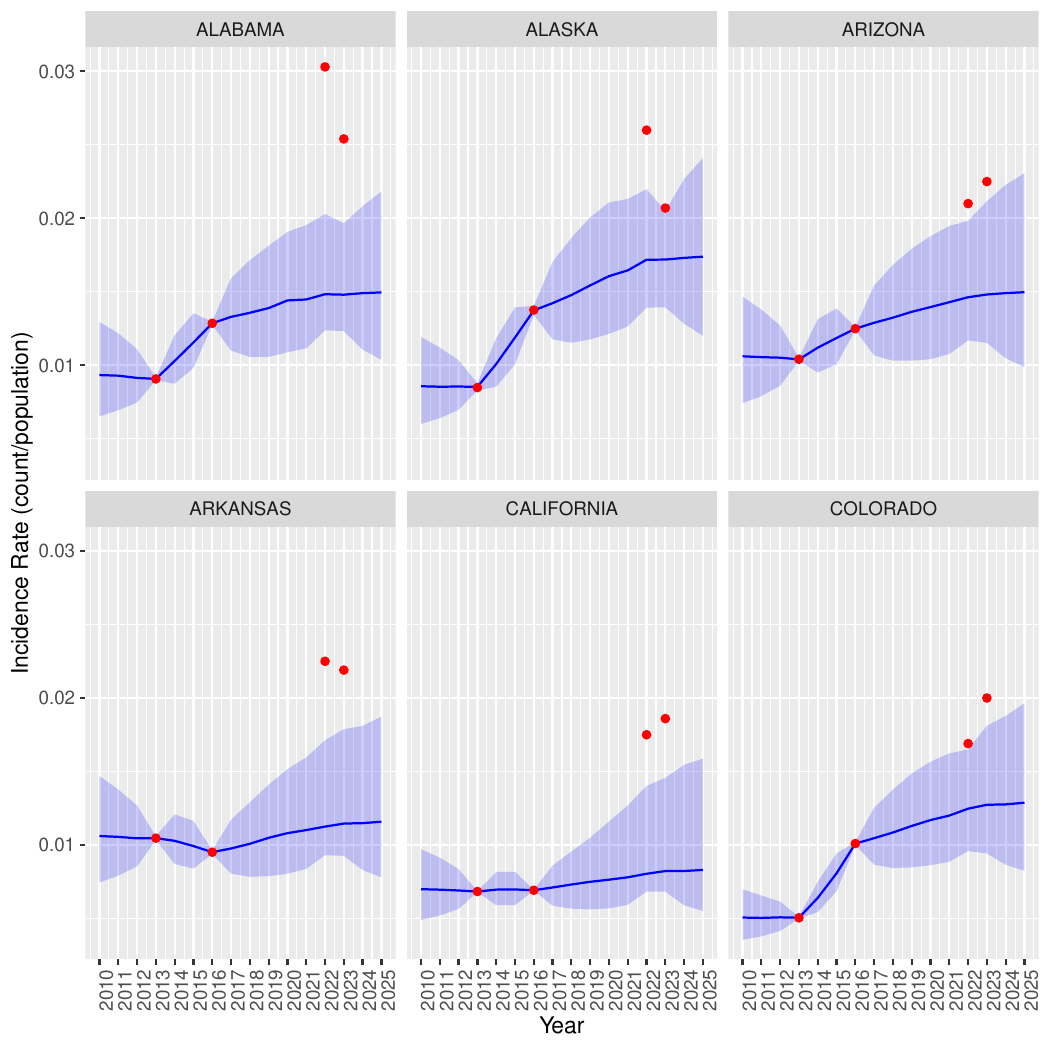}
  \label{fig:stateplots}
\end{figure}

\includepdf[pages=2-, pagecommand={}, scale=0.95]{state_plots.pdf}

\section{County Trends in OUD}\label{sec:countyfull}
\begin{figure}[H]
  \centering
  \includegraphics[width=\textwidth]{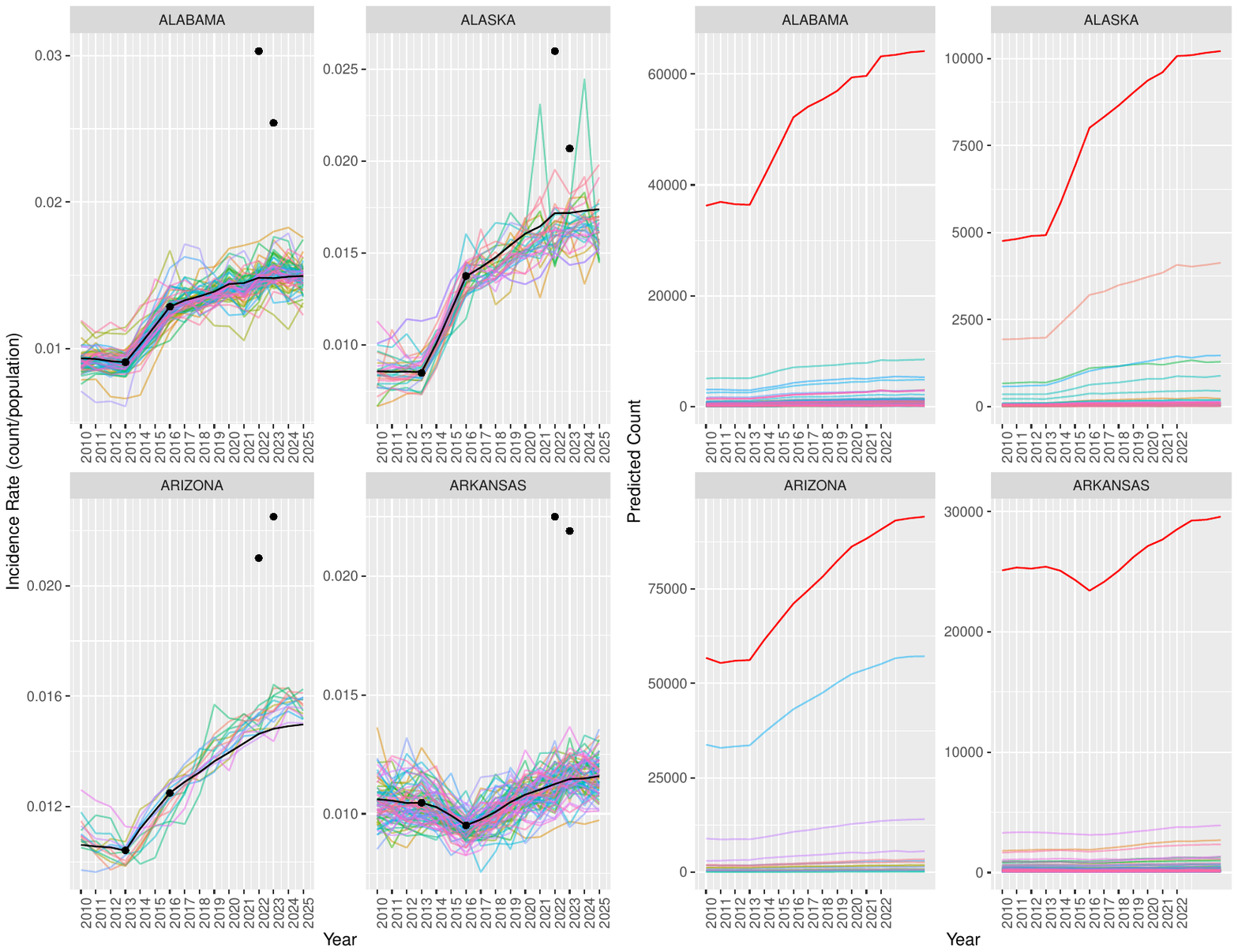}
  \label{fig:countyplots}
\end{figure}

\includepdf[pages=2-, pagecommand={}, scale=0.95]{county_plots.pdf}

\section{County Trends in OUD for Discontinuous Counties}\label{sec:discs}
\begin{figure}[ht]
  \centering
  \includegraphics[width=0.9\textwidth]{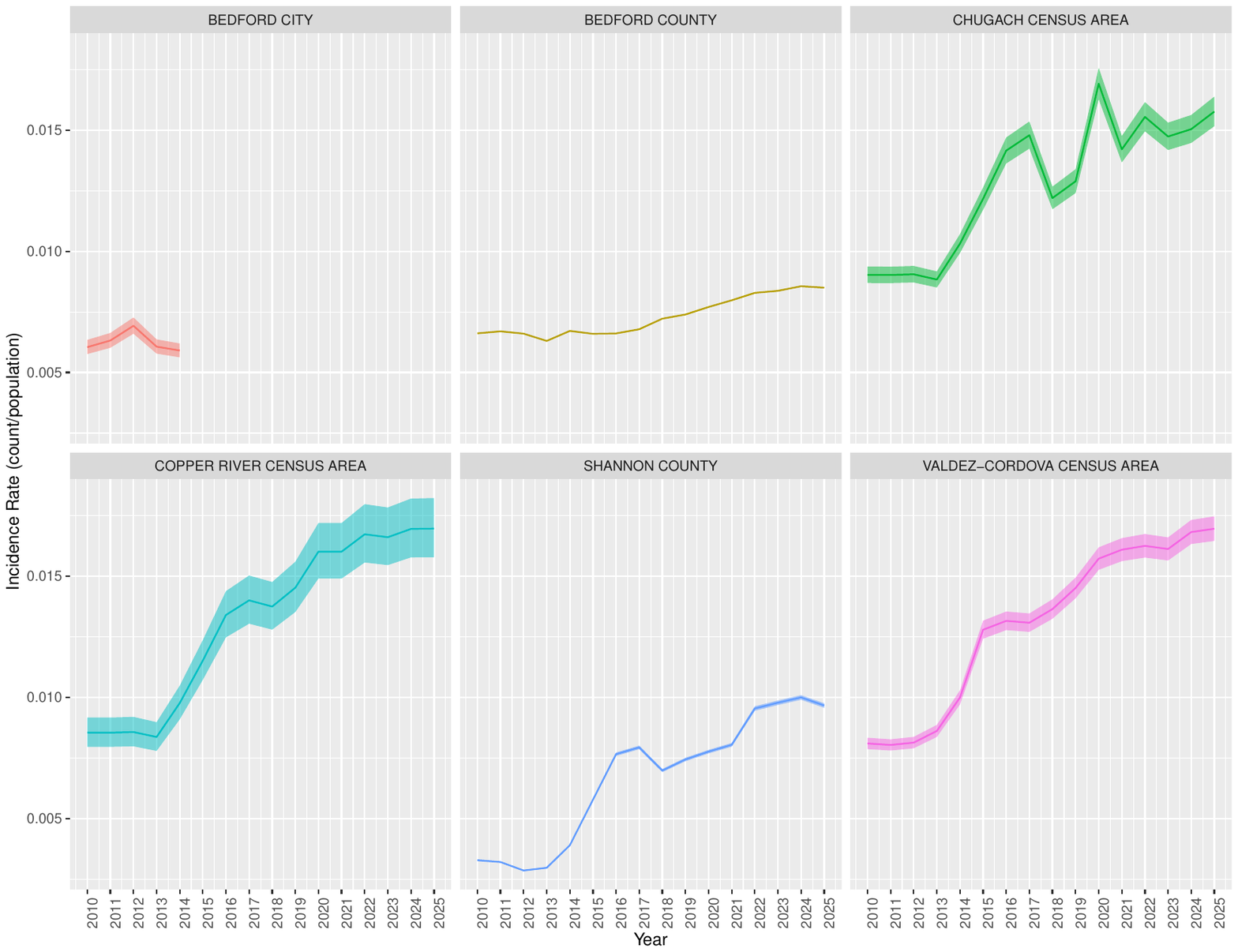}
  \caption{County level trends in OUD for counties that are discontinuous }
  \label{fig:discs}
\end{figure}
 
 \end{document}